\documentclass[11pt]{article}
\usepackage{multirow}
\usepackage{amsmath,verbatim,amssymb}
\usepackage{color}
\usepackage{booktabs}
\usepackage{threeparttable}
\usepackage{natbib}
\usepackage{setspace}
\usepackage{ulem}
\usepackage{graphicx}
\newcommand{\wh}{\widehat}
\newcommand{\wt}{\widetilde}

\newcommand{\var}{\mbox{Var}}
\newcommand{\diag}{\mbox{diag}}

\newcommand{\OP}{O_p}
\newcommand{\op}{o_p}
\newcommand{\rmT}{{\rm{T}}}
\newcommand{\R}{\mathbb{R}}
\newcommand{\E}{\mathbb{E}}
\newcommand{\bzero}{\textbf{0}}
\newcommand{\bone}{\textbf{1}}
\newcommand{\bx}{\textbf{x}}
\newcommand{\bX}{\textbf{X}}
\newcommand{\bY}{\textbf{Y}}
\newcommand{\bQ}{\textbf{Q}}
\newcommand{\bW}{\textbf{W}}
\newcommand{\bZ}{\textbf{Z}}
\newcommand{\bz}{\textbf{z}}
\newcommand{\bq}{\textbf{q}}

\newcommand{\bU}{\textbf{U}}
\newcommand{\bV}{\textbf{V}}
\newcommand{\bJ}{\textbf{J}}

\newcommand{\bB}{\textbf{B}}

\newcommand{\bbeta}{\boldsymbol{\beta}}

\newcommand{\bSigma}{\boldsymbol{\Sigma}}
\newcommand{\balpha}{\boldsymbol{\alpha}}
\newcommand{\bOmega}{\boldsymbol{\Omega}}

\newcommand{\btheta}{\boldsymbol{\theta}}

\newcommand{\bgamma}{\boldsymbol{\gamma}}

\newcommand{\bepsilon}{\boldsymbol{\epsilon}}
\newcommand{\bphi}{\boldsymbol{\phi}}

\newcommand{\bvarphi}{\boldsymbol{\varphi}}
\newcommand{\obsY}{T}
\newcommand{\failureT}{\wt{T}}
\newcommand{\Censore}{C}
\newcommand{\obsYa}{\obsY(a)}
\newcommand{\failureTa}{\failureT(a)}
\newcommand{\Censorea}{\Censore(a)}
\newcommand{\obsYTreat}{\obsY(1)}
\newcommand{\failureTTreat}{\failureT(1)}
\newcommand{\CensoreTreat}{\Censore(1)}
\newcommand{\obsYControl}{\obsY(0)}
\newcommand{\failureTControl}{\failureT(0)}
\newcommand{\CensoreControl}{\Censore(0)}

\newcommand{\CensoreIndicator}{\delta}
\newcommand{\predictorX}{\bX}

\newcommand{\Treat}{A}
\newcommand{\SetIndicator}{S}
\newcommand{\alldata}{\mathcal{OB}}
\newcommand{\individualdata}{\alldata_i}
\newcommand{\numRCT}{n_1}     
\newcommand{\numRWD}{n_0}
\newcommand{\numAll}{n}
\newcommand{\alldim}{p}
\newcommand{\HTE}{\tau}

\newcommand{\HTEbpara}{\balpha}
\newcommand{\HTEpara}{\alpha}
\newcommand{\UC}{u_c}
\newcommand{\UCbpara}{\bbeta}
\newcommand{\UCpara}{\beta}
\newcommand{\Allpara}{\theta}
\newcommand{\Allbpara}{\btheta}
\newcommand{\MeanOutcome}{\mu}
\newcommand{\propen}{e}
\newcommand{\nuisan}{\eta}
\newcommand{\AllX}{\bU}
\newcommand{\Allx}{U}
\newcommand{\XS}{\bZ}
\newcommand{\XSA}{\wt{\bZ}}
\newcommand{\tuneHTEfixed}{\gamma}
\newcommand{\MCP}{\rho}
\newcommand{\penaltydot}{\dot{\rho}} 
\newcommand{\penaltyddot}{\ddot{\rho}} 
\newcommand{\setUC}{\mathcal{G}} 
\newcommand{\tuneHTE}{\lambda_1}
\newcommand{\tuneUC}{\lambda_2}
\newcommand{\loss}{\ell}
\newcommand{\dMCP}{\penaltydot} 
\newcommand{\ddMCP}{\penaltyddot} 
\newcommand{\HTEbparahat}{\wh{\HTEbpara}}

\newcommand{\UCbparahat}{\wh{\UCbpara}}

\newcommand{\Allparahat}{\wh{\Allpara}}
\newcommand{\Allbparahat}{\wh{\Allbpara}}
\newcommand{\MeanOutcomehat}{\wh{\mu}}
\newcommand{\propenhat}{\wh{e}}
\newcommand{\nuisanhat}{\wh{\eta}}
\newcommand{\assAA}{\textbf{(A0)}}

\newcommand{\assM}{\textbf{(M0)}}
\newcommand{\assStute}{\textbf{(B0)}}
\newcommand{\assEigen}{\textbf{(B1)}}
\newcommand{\assRho}{\textbf{(B2)}}
\newcommand{\assRate}{\textbf{(B3)}}
\newcommand{\assG}{\textbf{(B4)}}

\newcommand{\assStuter}{\textbf{(C0)}}
\newcommand{\assEigenr}{\textbf{(C1)}}
\newcommand{\assRater}{\textbf{(C2)}}
\newcommand{\assGr}{\textbf{(C3)}}

\newcommand{\partI}{\text{(I)}}
\newcommand{\partIi}{\text{(I.1)}}
\newcommand{\partIii}{\text{(I.2)}}
\newcommand{\partIiii}{\text{(I.3)}}
\newcommand{\partII}{\text{(II)}}

\newcommand{\partIII}{\text{(III)}}
\newcommand{\partIIIi}{\text{(III.1)}}
\newcommand{\partIIIii}{\text{(III.2)}}
\newcommand{\partIV}{\text{(IV)}}
\newcommand{\Fc}{G}
\newcommand{\Fzt}{F^0}
\newcommand{\Fy}{H}
\newcommand{\Hend}{\tau_H}
\newcommand{\Ftild}{\wt{F}^0}
\newcommand{\setH}{\mathcal{H}}
\newcommand{\Htild}{\wt{H}}
\newcommand{\stuteGam}{\gamma}
\newcommand{\bstuteGam}{\bgamma}
\newcommand{\scorefb}{\bphi}
\newcommand{\scoref}{\phi}
\newcommand{\Fcr}{G_r}
\newcommand{\Fztr}{F^{0}_r}
\newcommand{\Fyr}{H_r}
\newcommand{\Hendr}{\tau_{H_r}}
\newcommand{\Ftildr}{\wt{F}^{0}_r}
\newcommand{\XA}{\wt{\predictorX}} 
\newcommand{\scorefbr}{\bvarphi}
\newcommand{\scorefr}{\varphi}

\newcommand{\rate}{R_{\numAll}}
\newcommand{\rater}{R_{\numRCT}}
\newcommand{\rateMu}{R_{\mu,\numAll}}
\newcommand{\ratePropen}{R_{e,\numAll}}
\newcommand{\realdim}{d_{\numAll}}
\newcommand{\realdimHTE}{d_{1\numAll}}
\newcommand{\realdimUC}{d_{2\numAll}}
\newcommand{\realdimr}{d_{1\numAll}}
\newcommand{\ratePropenr}{R_{e,\numRCT}}
\newcommand{\rateMur}{R_{\mu,\numRCT}}
\newcommand{\supp}{\mathcal{D}}

\newcommand{\setF}{\mathcal{F}}
\newcommand{\setALLbpara}{\Theta}
\newcommand{\envelopF}{F}
\newcommand{\setHTEbpara}{\Theta_r}

\newcommand{\bYhat}{\wh{\bY}}
\newcommand{\Lmb}{\Lambda}
\newcommand{\Lmbhat}{\wh{\Lmb}}

\newcommand{\dAllbpara}{\Delta\Allbpara}
\newcommand{\dAllpara}{\Delta\Allpara}

\newcommand{\Estute}{\wt{\E}_n}
\newcommand{\bqT}{\bq^\rmT}
\newcommand{\Gd}{D_*}
\newcommand{\bHessian}{\bJ_*}
\newcommand{\Gstute}{\wt{\mathbb{G}}_n}

\newcommand{\dnuisance}{\Delta\nuisan}

\newcommand{\propS}{r_{_S}}

\allowdisplaybreaks

\newtheorem{thm}{Theorem}[section]
\newtheorem{remark}{Remark}[section]

\title{Integrative Analysis of High-dimensional RCT and RWD Subject to Censoring and Hidden Confounding}

\author{Xin Ye$^{1,4}$, \quad Shu Yang$^{2}$\thanks{Corresponding author: syang24@ncsu.edu }, \quad Xiaofei Wang$^{3}$, \quad Yanyan Liu$^{4}$\\
	\small\it $^{1}$ School of Statistics and Mathematics, Guangdong University of Finance and Economics \\
	\small\it $^{2}$ Department of Statistics, North Carolina State University\\
	\small\it $^{3}$ Department of Biostatistics and Bioinformatics, Duke University \\
	\small\it $^{4}$ School of Mathematics and Statistics, Wuhan University.}
\date{}
	
\begin{document}
	
	\maketitle

\begin{abstract}	
	In this study, we focus on estimating the heterogeneous treatment effect (HTE) for survival outcome. The outcome is subject to censoring and the number of covariates is high-dimensional. We utilize data from both the randomized controlled trial (RCT), considered as the gold standard, and real-world data (RWD), possibly affected by hidden confounding factors. To achieve a more efficient HTE estimate, such integrative analysis requires great insight into the data generation mechanism, particularly the accurate characterization of unmeasured confounding effects/bias. With this aim, we propose a penalized-regression-based integrative approach that allows for the simultaneous estimation of parameters, selection of variables, and identification of the existence of unmeasured confounding effects. The consistency, asymptotic normality, and efficiency gains are rigorously established for the proposed estimate. 
	Finally, we apply the proposed method to estimate the HTE of lobar/sublobar resection on the survival of lung cancer patients. The RCT is a multicenter non-inferiority randomized phase 3 trial, and the RWD comes from a clinical oncology cancer registry in the United States. The analysis reveals that the unmeasured confounding exists and the integrative approach does enhance the efficiency for the HTE estimation.	
	
	\vspace{0.1in}
	
	
\end{abstract}

\section{Introduction}
\label{intro}
Recently, there has been a growing focus on the heterogeneity of treatment effect (HTE), a vital path towards personalized medicine (\citealp{Hamburg2010,Collins2015}). Accommodating confounding effects is crucial for for obtaining well-estimated HTE.
In such comparative medical research, it is important but challenging to fully determine what causes confounding effects and measure all of them. The most common approach is to conduct randomized controlled trials (RCTs). RCTs are known as the gold standard for assessing the causal effect of an intervention or treatment on the outcome of interest. The randomization allows the distribution of covariates in different groups to be balanced. However, RCTs have major downsides. For instance, they are costly and time-consuming, and often an inadequate sample size may result from recruitment challenges.

On the other hand, the increasing availability of real-world data (RWD) for research purposes, including electronic health records and disease registries, offers a broader demographic and diversity than RCTs. 
RWD provides abundant additional evidence to support HTE. Under the assumption that the records in RWD contain all the confounders, many approaches to harmonizing evidence from RCTs and RWD for HTE estimation have been developed, ranging from classic methods such as regression-based and inverse probability weighting to more recent machine learning models like neural networks (\citealp{Shalit2017}) and random forests (\citealp{Wager2018}). 
Inspired by Robinson transform (\citealp{Robinson1988}), \citet{Nie2018} recently proposed an R-learner to estimate HTE. The R-learner possesses the property of Neyman orthogonality (\citealp{Neyman1959}), enabling the integration of more extensive and flexible machine-learning methods for estimating the nuisance functions. However, it is always possible that in uncontrolled real-world settings, important confounders may be overlooked or unmeasured. For instance, doctors assign treatment based on patient's symptoms that are not documented in the medical chart. Unmeasured confounding can lead to unidentifiable causal effects of interest and result in distorted estimates of HTE. 

Classical approaches, such as instrumental variable methods (\citealp{Angrist1996}), negative controls (\citealp{Kuroki2014}), and sensitivity analysis (\citealp{Robins1999}),  have been proposed to address biases caused by hidden confounding. In recent years, a promising strategy to overcome the challenges posed by hidden confounding is to characterize the confounding function in RWD, and then utilize RCTs to identify both the HTE and confounding function. Drawing upon this idea, \citet{Kallus2018} proposed a regression-based method to estimate HTE. 
\citet{Yang2020} established the semiparametric efficient score function to estimate the HTE and confounding function  and demonstrated that their method can not only address issues arising from hidden confounding but also enhance the efficiency of HTE estimates.
Additionally, they introduced a testing procedure to ascertain the presence of unmeasured confounding, which informs the decision on whether to integrate RWD for a joint analysis (\citealp{YangTest2020}). However, once unmeasured confounding is detected, their approach discards all RWD data.
More recently, \citet{Wu22} leveraged the benefits of the R-learner to develop an integrative method for estimating the HTE and confounding function, utilizing experimental data for model identification and observational data for efficiency boosting.

However, the approaches mentioned above are all limited to fully observed data and low-dimensional covariates.
With the ongoing advancements in data acquisition technology and cloud storage, there is a growing trend towards the collection of high-dimensional data.
Censoring frequently occurs in various fields, especially for survival data, where the exact time of the event of interest cannot be observed due to the limited duration of the study.
Literature on estimating HTE from high-dimensional or censored data typically assumes the ideal case with no hidden confounding.
\citet{Ma2018} proposed characterizing the hazard ratio to mimic the heterogeneous treatment effect, yet they did not take into account high-dimensional covariates.
\citet{Zhu2020} used the difference in survival functions to describe heterogeneous treatment effects.
\citet{Hu2021} utilized the difference in survival quantile to characterize the survival treatment effect at the individual level and adopted a machine learning approach for model estimation.
\citet{Zhou2021} applied the sufficient dimension reduction technique to high-dimensional data without censoring.
To our knowledge, the literature on estimating HTE from high-dimensional censored data while offsetting the unmeasured confounding effect remains scarce. 

In this paper, we focus on improving the estimate of HTE for a survival outcome by integrating high-dimensional censored RCTs and RWD data, particularly in situations where unmeasured confounding may exist. We propose an integrative regression approach to simultaneously estimate parameters, select important variables, and determine the presence of unmeasured confounding effects. 
The proposed method assumes the transportability of the HTE. Therefore, the RCTs can be utilized to identify the HTE in RWD. Both the HTE and confounding function can be estimated through regularized weighted least square regression to accommodate censoring. The proposed method possesses the property of Neyman orthogonality, making it possible to adopt flexible machine-learning methods for the estimations of the nuisance functions.
are rigorously established, including estimation consistency, variable selection consistency, and asymptotic normality.
We demonstrate that the proposed integrative method results in a more efficient HTE estimate, at least on par with estimates solely based on RCTs data. When there is unmeasured confounding, instead of excluding all data from RWD, the proposed method can still make use of the RWD data in some cases. 
This study has the potential to enhance the existing literature in multiple important aspects.
First, an integrative analysis to include high-dimensional censored RWD data in HTE estimation is conducted, which can be more challenging than analyzing low-dimensional completely observable data.
Secondly, the proposed approach permits the presence of unmeasured confounding, which is more flexible and complements the analysis that assumes the unconfoundness in RWD. 
Thirdly, the proposed approach can identify whether the unmeasured confounding effect exists in a fully data-driven manner. This can contribute to more accurate estimates and lead to a deeper understanding of the data generation mechanism. Lastly, and equally importantly, this study offers a valuable practical tool for addressing a wide range of scientific issues. In particular, we apply the proposed integrative approach to improve the estimate of HTE on overall survival for patients with early-stage non-small-cell lung cancer undergoing lobar resection and limited resection, which convincingly demonstrates the usefulness of the proposed method. 

The remaining part of the paper is organized as follows. In Section 2, we introduce the proposed method. Theoretical properties are provided in Section 3. Numerical studies are conducted in Section 4, and application to real data is presented in Section 5. Concluding remarks are given in Section 6. Technical details are given in the Appendix.

\section{Methods}

Let $\failureT$ be the failure time, $\Censore$ be the censored time, $\obsY=\min(\failureT,\Censore)$ be the observation with censoring indicator $\CensoreIndicator=I(\failureT\leq \Censore)$, and $\Treat=0,1$ be the binary treatment variable. Let $\predictorX\in\R^p$ be the covariates vector, which includes the intercept term $X_0\equiv1$.
Let $\SetIndicator$ denote the data source, taking the value of 0 for RWD and 1 for RCT. The sample size of RCT is $\numRCT$ and RWD is $\numRWD$. Let the observed data be $\alldata=\left\{\individualdata,i=1,2,...,\numAll=\numRCT+\numRWD\right\}$, where
$\individualdata=(\obsY_i,\CensoreIndicator_i,\predictorX_i,\Treat_i,\SetIndicator_i).$
Under the potential outcome framework, denote that $\failureTa$, $\Censorea$ and $\obsYa=\min\left\{\failureTa,\Censorea\right\}$ be the potential failure time, potential censored time and potential observed time under treatment $a \in\left\{0,1\right\}$ respectively. We aim to evaluate the heterogeneous treatment effect (HTE) defined as follows
$$\HTE(\predictorX)=\E\left\{\log\left(\failureTTreat\right)-\log\left(\failureTControl\right)|\predictorX\right\}.$$
The definition of the HTE aligns seamlessly with conventional survival models, as illustrated, e.g., in (\ref{M0}) and (\ref{M1}).
The basic assumptions for modelling are as follows:

\begin{itemize}
	\item [$\assAA$]
	
	(i) $\failureT=\Treat\failureTTreat+(1-\Treat)\failureTControl$, $\Censore=\Treat\CensoreTreat+(1-\Treat)\CensoreControl$, and $\obsY=\Treat\obsYTreat+(1-\Treat)\obsYControl$.
	
	
	(ii) $\failureTa\perp\Treat|(\predictorX,\SetIndicator=1)$, $a\in\left\{0,1\right\}$.
	
	(iii)  $\E\left\{\log\left(\failureTTreat\right)-\log\left(\failureTControl\right)|\predictorX\right\}=\E\left\{\log\left(\failureTTreat\right)-\log\left(\failureTControl\right)|\predictorX,\SetIndicator\right\}$.
\end{itemize}

\begin{remark}
	(i) assumes that the consistency between observation and potential outcome holds.
	(ii) holds for the RCT by default. (iii)  states that the HTE is the same for the trial participants and the patient population at large.  It holds that if trial participants are randomly recruited for each subgroup of X, or the exclusion criteria of trial participation do not affect the treatment response.
\end{remark}

Define $\MeanOutcome_a(\predictorX,\SetIndicator)=\E\left\{\log(\failureT)|\Treat=a,\SetIndicator,\predictorX\right\}$, $a=0,1$. 
By assumption, it can be seen that for RCT, $\MeanOutcome_1(\predictorX,\SetIndicator=1)-\MeanOutcome_0(\predictorX,\SetIndicator=1)=\HTE(\predictorX)$. However, this equation may not hold in RWD if unmeasured confounding exists.
Define the confounding function $\UC(\predictorX)=\MeanOutcome_1(\predictorX,\SetIndicator=0)-\MeanOutcome_0(\predictorX,\SetIndicator=0)-\HTE(\predictorX)$. It can be seen that $\UC(\predictorX)$ captures the unmeasured confounding effect. The above formulations can be summarized into
$$
\MeanOutcome_1(\predictorX,\SetIndicator)-\MeanOutcome_0(\predictorX,\SetIndicator)=\HTE(\predictorX)+(1-\SetIndicator)\UC(\predictorX).
$$
By this formulation, we assume the following model on the failure time
\begin{equation}\label{M0}
	\log\failureT=\MeanOutcome_0(\predictorX,\SetIndicator)  +\Treat\HTE(\predictorX)+\Treat(1-\SetIndicator)\UC(\predictorX)+\epsilon,
\end{equation}
where $\E(\epsilon|\predictorX,\Treat,\SetIndicator)=0$, and $\E(\epsilon^2|\predictorX,\Treat,\SetIndicator)$ is finite. 
Taking expectation conditional on $(\predictorX,\SetIndicator)$ on both sides of this model leads
\begin{equation}\label{M1}
	\E\left\{\log(\failureT)|\SetIndicator,\predictorX\right\}=\MeanOutcome_0(\predictorX,\SetIndicator) +\propen(\predictorX,\SetIndicator)\HTE(\predictorX)+\propen(\predictorX,\SetIndicator)(1-\SetIndicator)\UC(\predictorX),
\end{equation}
where $\propen(\predictorX,\SetIndicator)=\E(A|\predictorX,\SetIndicator)$ is the propensity score. 
Calculating (\ref{M0}) minus  (\ref{M1}) leads to
\begin{equation}\label{BasicModel}
	\log(\failureT)=\MeanOutcome(\predictorX,\SetIndicator)  +\left\{\Treat-\propen(\predictorX,\SetIndicator)\right\}\HTE(\predictorX)+\left\{\Treat-\propen(\predictorX,\SetIndicator)\right\}(1-\SetIndicator)\UC(\predictorX)+\epsilon,
\end{equation}
where $\MeanOutcome(\predictorX,\SetIndicator)=\E\left\{\log(\failureT)|\SetIndicator,\predictorX\right\}$, $\E(\epsilon|\predictorX,\Treat,\SetIndicator)=0$, and $\E(\epsilon^2|\predictorX,\Treat,\SetIndicator)<\infty$.
Based on assumption $\assAA$, the above-induced formulation (\ref{BasicModel}) is an accelerated failure time (AFT) model. AFT model is a natural choice for clinical decision-making, because it has an intuitive regression interpretation on failure time. There is rich literature considering the AFT model for observational studies (\citealp{Henderson2020,Hu2021,Simoneau2020,YangAFT2020}).
Estimation of the AFT model with an unspecified error distribution has been studied extensively. Here, we adopt the weighted least squares (LS) approach (\citealp{Stute1993}) which is computationally more feasible. 

\begin{remark}
	More generally, instead of a logarithmic transformation on failure time, any other known monotone transformation can be considered. Then, the definition of HTE and assumption $\assAA$ should be correspondingly modified. 
\end{remark}

In (\ref{BasicModel}), we aim to estimate the HTE $\HTE$ and the confounding function $\UC$, with $\propen$ and $\MeanOutcome$ being the nuisance functions. 
First, we make the following assumptions for modelling heterogeneous treatment effects and unmeasured confounding effects.

\begin{itemize}
	\item [$\assM$] $\UC(\predictorX)$ can be modelled by $\predictorX^\rmT\UCbpara$, and $\HTE(\predictorX)$ can be modelled by $\predictorX^\rmT\HTEbpara$, where $\UCbpara$, $\HTEbpara\in\R^p$. 
\end{itemize}

Define the parameters of interests be $\Allbpara=(\HTEbpara^\rmT,\UCbpara^\rmT)^\rmT$; nuisance functions be $\nuisan=(\propen,\MeanOutcome)$. Let $\XS=(\predictorX,\SetIndicator)$, and $\AllX=(\predictorX,(1-\SetIndicator)\predictorX)$.
Then, the weighted loss function is
\begin{equation}\label{loss}
	\loss(\Allbpara,\nuisan|\alldata)=\sum_{i=1}^\numAll w_{i}\left[\log(\obsY_{(i)})-\MeanOutcome(\XS_{(i)})-\left\{\Treat_{(i)}-\propen(\XS_{(i)})\right\}\AllX_{(i)}^\rmT\Allbpara\right]^2,
\end{equation}
where $\obsY_{(1)}\leq \obsY_{(2)}\leq...\leq \obsY_{(\numAll)}$, and $\XS_{(i)}$, $\AllX_{(i)}$ are in corresponding order, $w_{i}$ is defined as follows
$$
w_{1}=\frac{\CensoreIndicator_{(1)}}{\numAll},\quad w_{i}=\frac{\CensoreIndicator_{(i)}}{\numAll-i+1}\prod_{j=1}^{i-1}\left(\frac{\numAll-j}{\numAll-j+1}\right)^{\CensoreIndicator_{(j)}},\quad i=2,3,...,\numAll.
$$
Suppose that the nuisance function $\nuisan$ can be pre-estimated, then we propose to use the following penalized regression to get the estimate. It can simultaneously select important variables and determine whether the unmeasured confounding exists:
\begin{equation}\label{ploss}			\loss_{\tuneHTE,\tuneUC}(\Allbpara,\nuisanhat|\alldata)=\loss(\Allbpara,\nuisanhat|\alldata)		+\sum_{j=1}^{\alldim}\MCP(\vert\Allpara_j\vert;\tuneHTE)+\sum_{j=\alldim+1}^{2\alldim}\MCP(\vert\Allpara_j\vert;\tuneUC).
\end{equation}
where $\MCP(t;\lambda)$ is a penalty function with tuning parameters $\lambda>0$ to recover sparsity, various kinds of penalty functions can be used to derive sparse and unbiased estimates, such as adaptive Lasso (\citealp{Zou2006}), SCAD (\citealp{Fan2004}), and MCP (\citealp{Zhang2010}).
It can be seen that the penalty function in (\ref{ploss}) consists of two parts with tuning parameters $\tuneHTE$, $\tuneUC$ respectively. The first part corresponds with the parameter of HTE, i.e., $\HTEbpara$, and the second part corresponds with the parameter of unmeasured confoundings, i.e., $\UCbpara$. 
By adopting penalties respectively, the method can fit in with a more general case where the sizes of coefficient in HTE and confounding function are different. A Similar strategy can be found in \citet{Cheng2023}.
The final estimate can be written as
\begin{equation}\label{finalEstimate}
	\Allbparahat=\arg\min_{\Allbpara}\loss_{\tuneHTE,\tuneUC}(\Allbpara,\nuisanhat|\alldata),
\end{equation}
where the tuning parameters $\tuneHTE$ and $\tuneUC$ can be selected by criteria such as AIC, BIC, and cross-validation (CV).

\section{Theoretical properties}
Denote that the true parameters be $\Allbpara^*=(\balpha^{*\rmT},\bbeta^{*\rmT})^\rmT$, and true nuisance functions be $\nuisan^*$. Define index sets of non-zero parameters as follows:  $\supp=\left\{1\leq j\leq 2\alldim|\Allpara^*_j\neq 0\right\}$ with element number $\realdim$, $\supp_1=\left\{1\leq j\leq \alldim|\alpha^*_j\neq 0\right\}$ with element number $\realdimHTE$, $\supp_2=\left\{1\leq j\leq \alldim|\beta^*_j\neq 0\right\}$ with element number $\realdimUC$.
Following the notations in \citet{Stute1996}, let $\Fc$ be the probability distribution function (p.d.f) of $\Censore$, with $\tau_{G}=\inf\left\{x:G(x)=1\right\}$, $F$ be the p.d.f of $\failureT$, with $\tau_{F}=\inf\left\{x:F(x)=1\right\}$, and
$\Fy$ be the p.d.f of $\obsY$, with $\Hend=\inf\left\{x:\Fy(x)=1\right\}$. 
Let $\Fzt\in\mathcal{P}$ be the p.d.f of $(\XSA,\failureT)$, where $\XSA=(\XS,\Treat)$. Define
\begin{equation*}
	\Ftild(\bz,t)=\left\{
	\begin{array}{lc}
		\Fzt(\bz,t), & t<\Hend,\\
		\Fzt(\bz,\Hend-)+\Fzt(\bz,\Hend)I(\Hend\in\setH),& t\geq\Hend,\\
	\end{array}
	\right.
\end{equation*}
with $\setH$ denoting the set of atoms of $\Fy$, possibly empty. 
Define the score function
$$
\scoref_j(\XSA,\failureT;\Allbpara^*,\nuisan^*)=\left\{A-\propen^*(\XS)\right\}\Allx_j\left[\log(\failureT)-\MeanOutcome^*(\XS)-\left\{A-\propen^*(\XS)\right\}\AllX^\rmT\Allbpara^*\right],
$$
where $j=1,2,...,2\alldim$.
It can be seen that $\E\scorefb(\XSA,\failureT;\Allbpara^*,\nuisan^*)=\bzero$, where $\scorefb=\left(\scoref_j,j=1,2,...,2\alldim\right)$. 
Define $\stuteGam_0(y)=\exp\left\{\int^{y-}_0\left\{1-\Fy(v)\right\}^{-1}\Htild^0(dv)\right\}$, where
	$\Htild^{0}(y)=\Pr(\obsY\leq y, \CensoreIndicator=0)$.

\begin{itemize}
	\vspace{0.1in}
	\item [\assStute] (i) $\Pr(\failureT\leq\Censore|\predictorX,\SetIndicator,\Treat,\failureT)=\Pr(\failureT\leq\Censore|\failureT)$.
	For $j=1,2,...,2\alldim$, $\Fzt\in\mathcal{P}$, 
	
	(ii) The p.d.f. $F$ and $G$ have no jump in common, and $\tau_F<\tau_G$.
	
	(iii) $\E\left\{\scoref_j(\XSA,\obsY;\Allbpara^*,\nuisan^*)\stuteGam_0(\obsY)\CensoreIndicator\right\}^2<\infty$;
	
	(iv) Let $g(y)=\int_0^{y-}\left\{1-H(w)\right\}^{-1}\left\{1-G(w)\right\}^{-1}G(dw)$. It holds that
	$$\int|\scoref_j(\bz,w;\Allbpara^*,\nuisan^*)|\sqrt{g(w)}\Ftild(d\bz,dw)<\infty.$$
	
	\vspace{0.1in}
	\item [\assEigen]	
	(i) The eigenvalues of $\E\left[\left\{\Treat-\propen^*(\XS)\right\}^2\AllX\AllX^\rmT\right]$ are larger than a positive constant $c_1$.
	
	(ii) The eigenvalues of $\E\AllX\AllX^\rmT$ are smaller than a positive constant $c_2$.

	\vspace{0.1in}
	\item [\assRho] The penalty function satisfies the following properties.
	
	(i) $\rho(x;\lambda)$ is nondecreasing in $x\in[0,\infty)$ and $\rho(0;\lambda)=0$.
	
	(ii) Let $\dMCP(x;\lambda)=\partial\MCP(x;\lambda)/\partial x$. It exists and is bounded in $x\in(0,\infty)$. 	 
	In addition,  $\dot{\rho}(x;\lambda)/\lambda>0$, as $x\rightarrow 0+$, $n\rightarrow\infty$, and $|\dMCP(x_1;\lambda)-\dMCP(x_2;\lambda)|\leq O(1)\lambda|x_1-x_2|$, for $x_1,x_2\in(0,\infty)$.
	
	(iii) Let $\ddMCP(x;\lambda)=\partial^2\MCP(x;\lambda)/\partial x^2$. It exists and is bounded in $x\in(\gamma_1\lambda,\infty)$, where $\gamma_1>0$ is a constant. It holds that $|\ddMCP(x_1;\lambda)-\ddMCP(x_2;\lambda)|\leq O(1)|x_1-x_2|$, for $x_1,x_2\in(\gamma_1\lambda,\infty)$.

	\vspace{0.1in}
	\item [\assRate] The pre-estimated nuisance parameter $\nuisanhat$ is independent of the samples used to build the loss function. Considering the abuse of notation, we continue to use $n$ to denote the sample size for constructing the loss function and assume that the pre-estimated nuisance parameter $\hat\eta$ is obtained from another sample of size $r_\eta n$, where $r_\eta n$ is bounded by a positive constant.
	Let $\ratePropen$ be the convergence rate of $\Vert\propenhat-\propen^*\Vert_\infty$, $\rateMu$ be the convergence rate of $\Vert\MeanOutcomehat-\MeanOutcome^*\Vert_\infty$. Define rate
	$\rate=\max\left\{\rateMu,\ratePropen\Vert\Allbpara^*\Vert_2,\sqrt{\alldim/\numAll}\right\}$. The real parameter satisfies $\min_{j\in\supp_1}|\HTEpara_j^*|/\tuneHTE\rightarrow\infty$, $\min_{j\in\supp_2}|\UCpara_j^*|/\tuneUC\rightarrow\infty$, as $n\rightarrow\infty$. Additional conditions are listed in the followings. 
	
	(i) $\mathop{\max}\limits_{j\in\supp_1}\left\{\dMCP(|\HTEpara_j^*|;\tuneHTE)\right\}=O(\rate/\sqrt{\realdimHTE})$, $\mathop{\max}\limits_{j\in\supp_2}\left\{\dMCP(|\UCpara_j^*|;\tuneUC)\right\}=O(\rate/\sqrt{\realdimUC})$.
	
	(ii) $\mathop{\max}\limits_{j\in\supp_1}\left\{|\ddMCP(|\HTEpara_j^*|;\tuneHTE)|\right\}=o(1)$, $\mathop{\max}\limits_{j\in\supp_2}\left\{|\ddMCP(|\UCpara_j^*|;\tuneUC)|\right\}=o(1)$.
	
	(iii) $\mathop{\max}\limits_{j\in\supp_1}\left\{\dMCP(|\HTEpara_j^*|;\tuneHTE)\right\}=O(1/\sqrt{\numAll\realdimHTE})$,  $\mathop{\max}\limits_{j\in\supp_2}\left\{\dMCP(|\UCpara_j^*|;\tuneUC) \right\}=O(1/\sqrt{\numAll\realdimUC})$.
	
	(iv)$\sqrt{\numAll}\rate^2=o(1)$, and  $\sqrt{\numAll}\max\left\{\tuneHTE,\tuneUC\right\}\rate=o(1)$.
	
	(v)
	$\E|\scoref_j(\XSA,\failureT;\Allbpara,\nuisan)-\scoref_j(\XSA,\failureT;\Allbpara^*,\nuisan^*)|^2\leq \left(\Vert\Allbpara-\Allbpara^*\Vert_2\vee\Vert\nuisan-\nuisan^*\Vert_\infty\right)^bc_3$, $j\in\supp$, where $b$ and $c_3$ are positive constants. In addition, $\sqrt{\realdim}\rate^{b/2}=o(1)$, $\sqrt{\realdim}n^{-1/2+1/q}=o(1)$, $q>2$.
	
	\vspace{0.1in}
	\item [\assG] 
	In what follows, we use $\Vert\cdot\Vert_{Q,q}$ to denote the $L^q(Q)$ norm. 
	The uniform entropy numbers for set $\mathcal{F}$ with radius $\xi>0$ under $L^q(Q)$ norm are defined as $\sup_Q \log N(\xi, \mathcal{F}, \Vert\cdot\Vert_{Q,q})$, where $N(\xi, \mathcal{F}, \Vert\cdot\Vert_{Q,q})$ is the corresponding covering number.
	Let $\setALLbpara=\left\{\Allbpara:\Vert\Allbpara-\Allbpara^*\Vert_2\leq \rate c_4\right\}$, where $c_4$ is a positive constant. Define class 
	$$
	\setF_{1,\nuisan}=\left\{\scoref_j(\cdot;\Allbpara,\nuisan):j\in\supp,\Allbpara\in\setALLbpara\right\},
	$$
	with measurable envelop $\envelopF_{1,\nuisan}$. It satisfies  $\Vert\envelopF_{1,\nuisan}\Vert_{\Fzt,q}\leq c_5$ where $c_5$ is a positive constant and $\Fzt\in\mathcal{P}$. It holds that for all $0<\xi\leq 1$, the uniform entropy number of $\setF_{1,\nuisan}$ obeys
	$$
	\sup_{Q\in\mathcal{P}}\log N(\xi\Vert\envelopF_{1,\nuisan}\Vert_{Q,2},\setF_{1,\nuisan},\Vert\cdot\Vert_{Q,2})\leq v\log\frac{c_6}{\xi},
	$$
	where $c_6$ is a positive constant.
\end{itemize}
\begin{remark}
	$\assStute$ guarantees that Stute's empirical probability measure converges to $\Fzt$ (\citealp{Stute1993}). In addition, it assures the asymptotic normality of $\sum_{i=1}^nw_{i}\scoref_j(\wt{\bZ}_{(i)},T_{(i)})$ given real nuisance functions (\citealp{Stute1996}). 	
	$\assStute$(i) assumes that the censoring variable is conditionally independent of $(\bX,A,S)$ given the failure time $\wt{T}$.
	By contrast, the utilization of the Inverse Censoring Probability Weight (IPCW) of the form $\delta/\Pr(C>T|\bX,A,S,T)$ often requires $C\perp \wt{T}|S,A,\bX$ instead. While Stute's weights are fully non-parametric, using the IPCW requires  estimating of the survival function for the censoring time, $\Pr(C>t|\bX,A,S)$, which may introduce additional model assumptions. We discuss the further development based on IPCW in Section 6.
	$\assEigen$ puts constraints on eigenvalues of design matrices. 
	$\assRho$ states the basic properties of the penalty function.
	Many penalty functions, such as SCAD and MCP, can meet these properties.
	$\assRate$ contains several assumptions on nuisance function, penalty function and convergence rate. 
	It should be noted that the assumption of independent pre-estimated nuisance parameter can be reached by data splitting. 
	$\assRate$(i)-(iii) naturally hold when the signal of the real parameter is strong enough. $\assRate$(iv) requires the convergence rate of the nuicance function estimate to be at least faster than $n^{-1/4}$, and $p=o(n^{1/2})$.
	$\assG$ is used to reach the condition in Lemma 6.2 in \citet{Chernozhukov2018}. 
\end{remark}

\begin{thm}[Consistency]\label{thm1}
	If $\assM$, $\assAA$, $\assStute$, $\assEigen$, $\assRho$ and $\assRate$ (i)(ii) hold, then $\Vert\Allbparahat-\Allbpara^*\Vert_2=\OP(\rate)$, where $\rate$ is defined in \assRate.
\end{thm}
\begin{thm}[Sparsity recovery]\label{thm2}
	Suppose the result in Theorem \ref{thm1} holds. If $\min\left\{\tuneHTE,\tuneUC\right\}\rate^{-1}\rightarrow \infty$ as $\numAll\rightarrow\infty$, then $\Pr\left(\Allbparahat_{\supp^c}=\bzero \right)\rightarrow 1$.
\end{thm}

First, we introduce some notations to derive the asymptotic normality of the proposed estimator.
Let 
	$\Htild^{1}(\bz,y)=\Pr(\XSA\leq \bz, \obsY\leq y, \CensoreIndicator=1)$, 
$$
\stuteGam_{1j}(y)=\frac{1}{1-\Fy(y)}\int 1_{\left\{y<v\right\}}\phi_j(\bz,v;\Allbpara^*,\nuisan^*)\stuteGam_0(v)\Htild^{1}(d\bz,dv),
$$
$$
\stuteGam_{2j}(y)=\int\int \frac{1_{\left\{v<y,v<w\right\}}\phi_j(\bz,w;\Allbpara^*,\nuisan^*)\stuteGam_0(v)}{\left\{1-\Fy(v)\right\}^2}\Htild^0(dv)\Htild^{1}(d\bz,dw),
$$
and $\bstuteGam_{1}(y)=\left\{\stuteGam_{1j}(y),j=1,2,...,2\alldim\right\}$, and $\bstuteGam_{2}(y)=\left\{\stuteGam_{2j}(y),j=1,2,...,2\alldim\right\}$. 
Define $\bV=\bB^{-1}\bSigma_{\supp}\bB^{-1}$, where $$\bB=\E\left[\left\{\Treat-\propen^*(\XS)\right\}^2\AllX_{\supp}\AllX_{\supp}^\rmT\right],$$  $$\bSigma=\var\left\{\scorefb(\XSA,\failureT;\Allbpara^*,\nuisan^*)\stuteGam_0(\failureT)\CensoreIndicator+\bstuteGam_1(\failureT)(1-\CensoreIndicator)-\bstuteGam_2(\failureT)\right\}.$$
\begin{thm}[Asymptotic normality]\label{thm3}
	Suppose the result of consistency and sparsity recovery hold. 
	Assume that $\assM$, $\assAA$, $\assStute$-$\assRho$ and $\assRate$ (iii)(iv)(v) hold. 
	For any $\bq\in\R^{\realdim}$, $\Vert\bq\Vert_2<\infty$, if $\sigma^2\bqT\bV\bq\rightarrow \sigma_*^2$ as $\numAll\rightarrow\infty$, then 
	$$
	\sqrt{n}\bqT\left(\Allbparahat_{\supp}-\Allbpara^*_{\supp}\right)\rightarrow_d N(0,\sigma_*^2).
	$$	
\end{thm}

We need additional conditions to derive the asymptotic properties of the RCT-only estimator. Since the conditions are similar to $\assStute$-$\assG$, details are presented in the Appendix. Define rate
$\rater=\max\left\{\rateMur,\ratePropenr\Vert\HTEbpara^*\Vert_2,\sqrt{\alldim/\numRCT}\right\}$, and
$\bV_r=\bB_r^{-1}\bSigma_{r\supp_1}\bB_r^{-1}$,
where
$$\bB_r=\E\left[\left\{\Treat-\propen^*(\XS)\right\}^2\predictorX_{\supp_1}\predictorX_{\supp_1}^\rmT|\SetIndicator=1\right],$$ $$\bSigma_r=\var\left\{\scorefbr(\XA,\failureT;\HTEbpara^*,\nuisan^*)\stuteGam_{r0}(\failureT)\CensoreIndicator+\bstuteGam_{r1}(\failureT)(1-\CensoreIndicator)-\bstuteGam_{r2}(\failureT)|\SetIndicator=1\right\}.$$
The definition of $\scorefbr$, $\stuteGam_{r0}$, $\bstuteGam_{r1}$, $\bstuteGam_{r2}$ are presented in Appendix for details.
\begin{thm}[Asymptotic normality for RCT-only estimator]\label{thm4}
	Suppose the result of consistency with rate $\rater$ and sparsity recovery hold. 
	Assume that $\assM$, $\assAA$, $\assStuter$, $\assEigenr$ and $\assRater$ (in the Appendix) hold. 		
	For any $\bq\in\R^{\realdimr}$, $\Vert\bq\Vert_2<\infty$, if $\sigma^2\bqT\bV_r\bq\rightarrow \sigma_{r*}^2$ as $\numRCT\rightarrow\infty$, then 
	$$
	\sqrt{\numRCT}\bqT\left(\HTEbparahat^{rct}_{\supp_1}-\HTEbpara^*_{\supp_1}\right)\rightarrow_d N(0,\sigma_{r*}^2).
	$$	
\end{thm}

\begin{thm}[Efficiency gain]\label{thm5}
	Suppose the results in Theorem \ref{thm3} and \ref{thm4} hold. If there is no censoring, it can be seen that $\bSigma_{r\supp_1}=\bB_r$, and $\bSigma_{\supp}=\bB$. For any $\bq\in\R^{\realdimr}$, $\Vert\bq\Vert_2<\infty$, with probability converging to 1, we have
	\begin{equation}\label{variance}
		\var(\sqrt{\numAll}\bq^\rmT\HTEbparahat_{\supp_1})\leq\var(\sqrt{\numAll}\bq^\rmT\HTEbparahat^{rct}_{\supp_1}),
	\end{equation} 
	where the equality holds if and only if there exists a $\realdimUC\times\realdimHTE$ constant matrix $\bQ$, such that when $S=0$,  $\predictorX_{\supp_1}=\bQ^\rmT\predictorX_{\supp_2}$. Specially, when $\supp_1\subset\supp_2$, the equality holds. When ${\supp_2}=\emptyset$, under (B1)(i), the ``$<$" in (\ref{variance}) strictly holds.
	
\end{thm}
\begin{remark}
	Censoring leads to a more complicated form of variance, thus it is difficult to see the efficiency gain directly. 
	Let $\bB=(\bB_{11},\bB_{12};\bB_{21},\bB_{22})$ where
	$\bB_{11}=\E\left\{\Treat-\propen^*(\XS)\right\}^2\predictorX_{\supp_1}\predictorX_{\supp_1}^\rmT$, $\bB_{12}=\E(1-\SetIndicator)\left\{\Treat-\propen^*(\XS)\right\}^2\predictorX_{\supp_1}\predictorX_{\supp_2}^\rmT$, $\bB_{22}=\E(1-\SetIndicator)\left\{\Treat-\propen^*(\XS)\right\}^2\predictorX_{\supp_2}\predictorX_{\supp_2}^\rmT$.  Define
	$\bOmega_{11}=\left(\bB_{11}-\bB_{12}\bB_{22}^{-1}\bB_{12}^\rmT\right)^{-1}$. Let $\bSigma_{11}$ be the submatrix of $\bSigma$ with columns and rows corresponding to $\HTEbpara^*_{\supp_1}$, $\bSigma_{12}$ be the submatrix with columns corresponding to $\HTEbpara^*_{\supp_1}$ and rows corresponding to $\UCbpara^*_{\supp_2}$,  $\bSigma_{22}$ be the submatrix with columns and rows corresponding to $\UCbpara^*_{\supp_2}$. Let $\propS=\Pr(\SetIndicator=1)$, and $\Delta\bSigma_{11}=
	\bSigma_{11}-\bB_{12}\bB_{22}^{-1}\bSigma_{12}^\rmT-\bSigma_{12}\bB_{22}^{-1}\bB_{12}^\rmT+\bB_{12}\bB_{22}^{-1}\bSigma_{22}\bB_{22}^{-1}\bB_{12}^\rmT$.
	Generally, if
	$$
	\bB_r^{-1}\bSigma_r\bB_r^{-1}-\propS\bOmega_{11}\Delta\bSigma_{11}\bOmega_{11}\geq 0,
	$$
	that is, the matrix is semi-definite, then the variance of the proposed estimate $\sqrt{\numAll}\bqT\HTEbparahat_{\supp_1}$ will not larger than the RCT-only estimate $\sqrt{\numAll}\bqT\HTEbparahat_{\supp_1}^{rct}$. 
		Specially when $\supp_2=\emptyset$, $\bB_{11}=\bB_r$ and $\bSigma_{11}=\bSigma_r$, i.e.,  the distributions of $(A,\bX)$ and censoring in RCT and RWD are similar, then the variance of the proposed estimate $\sqrt{\numAll}\bqT\HTEbparahat_{\supp_1}$ will be rigorously smaller than that of the RCT-only estimate $\sqrt{\numAll}\bqT\HTEbparahat_{\supp_1}^{rct}$. 
\end{remark}

\begin{remark}[Variance estimation]\label{var}
	The theoretical variances obtained in Theorems \ref{thm3} and \ref{thm4} are not easy to estimate based on the formulations. Following \citet{Huang2006}, we estimate the variance using the nonparametric 0.632 bootstrap (\citealp{Efron1993}), in which approximately $0.632\numAll$ samples from the $\numAll$ observations are randomly selected without replacement. 
\end{remark}

\section{Simulation}
We conduct simulation studies to evaluate the performance of the proposed method including efficiency gain (compared with the estimators that only use RCT data), parameters estimation, variable selection and identification of unmeasured confounding. The data is generated from the following model
$$
\log(\failureT)= \mu_0(\bX,S) +\Treat\predictorX^\rmT\HTEbpara^*+{(1-S)}u+\epsilon,
$$
where  $\mu_0(\bX,S)=\sin(X_1)+0.2X_4^2-0.5\predictorX^\rmT\HTEbpara^*-0.5(1-\SetIndicator)\predictorX^\rmT\UCbpara^*$. Here
$\bX$ is observable, while $u$ is the unmeasured confounding effect. 
	We generate $n=2500$ samples from this model with the following distributions: 
	$S\sim Bernoulli(0.2)$, 
$\Treat\sim Bernoulli(0.5)$, $\bX|\Treat\sim N(0.2\Treat\times(\bone_{8},\bzero_{\alldim-8}),\bSigma)$, $u|\Treat\sim N(\Treat\predictorX^\rmT\UCbpara^*,\bSigma)$, and $\epsilon\sim N(0,1)$, where  $\bSigma=(0.3^{|i-j|},i,j=1,2,...,\alldim)$. 
Let  $\HTEbpara^*=Signal\times(\bone_{4},-\bone_{4},\bzero_{\alldim-8})$,  $\UCbpara^*=Signal\times(\bone_{2},-\bone_{2},\bzero_{\alldim-4})$, $Signal=2$, provided that unmeasured confounding effect exists, otherwise $\UCbpara^*=\bzero_{\alldim}$. 
The dimension of $\bX$ is considered to be $p\in\left\{20,50\right\}$,  the censored time $\log C\sim \text{Unif}[t_0,t_1]$, and $t_0$, $t_1$ adjust the censored rate to be around 20\% or 40\%. 
We adopt the MCP function as the penalty function, i.e., $\MCP(t;\lambda)=\lambda\int_{0}^{t}\big(1-x/(\tuneHTEfixed\lambda)\big)_+dx$.

\subsection{Finite-Sample studies}
For the proposed method, cross-validation and BIC to select the tuning parameters and refer to them as RL.cv and RL.bic, respectively. that ignores the unmeasured confounding effect and refers to it as RL.NAI.
In addition, we compare the proposed method with the following methods:
\begin{itemize}
	\item [] {\it Outcome-adjusted method}: define the adjusted outcome
	\begin{equation*}
		\failureT^{adjust}=\frac{\Treat\left\{\log(\failureT)-\MeanOutcome_1(\XS)\right\}}{\propen(\XS)}+\MeanOutcome_1(\XS)-\frac{(1-\Treat)\left\{\log(\failureT)-\MeanOutcome_0(\XS)\right\}}{1-\propen(\XS)}-\MeanOutcome_0(\XS).
	\end{equation*}
	Under assumption $\assAA$ and $\assM$, $\E\left(\failureT^{adjust}|\XS\right)=\predictorX^\rmT\HTEbpara+(1-\SetIndicator)\predictorX^\rmT\UCbpara$. Then we can build the penalized regression model based on this equation (similar to the construction of the proposed method). We use the same penalty function as the proposed method to identify unmeasured confounding effect and adopt CV and BIC to select the tuning parameters. This method is referred to as OA.cv and OA.bic respectively.
	
	\vspace{0.1in}
	\item [] {\it AFT model with $\mu_0$}:
	under assumption $\assAA$ and $\assM$, it holds that
	$$
	\E\left\{\log(\failureT)-\MeanOutcome_0(\XS)|\XS,\Treat=1\right\}=\predictorX^\rmT\HTEbpara+(1-\SetIndicator)\predictorX^\rmT\UCbpara.
	$$
	Then we can build the AFT model based on this equation. The estimation procedures are the same as the outcome-adjusted method. This method is referred to as GM0.cv and GM0.bic respectively.
	
	\vspace{0.1in}
	\item []{\it AFT model with $\mu_1$}:
	under assumption $\assAA$ and $\assM$, it holds that
	$$
	\E\left\{\MeanOutcome_1(\XS)-\log(\failureT)|\XS,\Treat=0\right\}=\predictorX^\rmT\HTEbpara+(1-\SetIndicator)\predictorX^\rmT\UCbpara.
	$$
	Then we can build the AFT model based on this equation. The estimation procedures are the same as the outcome-adjusted method. This method is referred to as GM1.cv and GM1.bic respectively.
	
	\vspace{0.1in}
	\item []{\it The meta estimates}: combine GM0.cv and GM1.cv (GM0.bic and GM1.bic) by weights of sample size. This method is referred to as Meta.cv and Meta.bic respectively. 
	
	\vspace{0.1in}
	\item []{\it AFT model with $\mu_0$, $\mu_1$}: under assumption $\assAA$ and $\assM$, it holds that
	$$
	\E\left\{\MeanOutcome_1(\XS)-\MeanOutcome_0(\XS)|\XS\right\}=\predictorX^\rmT\HTEbpara+(1-\SetIndicator)\predictorX^\rmT\UCbpara.
	$$
	Then we can build the AFT model based on this equation. The following procedures are the same as the outcome-adjusted method. This method is referred to as GM01.cv and GM01.bic respectively.
\end{itemize}
We calculate the RCT-only estimates for all these methods and use CV to select tuning parameters referred to as RL.RCT, OA.RCT, GM0.RCT, GM1.RCT, Meta.RCT, GM01.RCT, respectively. In addition, assuming that we correctly select the variables, we can calculate the oracle estimates referred to as RL.or, RL.NAIor, OA.or, GM0.or, GM1.or, Meta.or, GM01.or, RL.RCTor, OA.RCTor, GM0.RCTor, GM1.RCTor, Meta.RCTor, GM01.RCTor, respectively.

For the estimation of HTE's parameter, we use mean square error (MSE) to evaluate the performance (variance) of estimation and use false discovery rate (FDR) to evaluate the performance of variable selection. 
The definitions are as follow: for simulation times $b=1,2,...,B$,
$\text{RMSE}=(\text{MSE})^{1/2}$, where
$
\text{MSE}=(Bp)^{-1}\sum_{b=1}^{B}\sum_{j=1}^{p}(\wh{\alpha}_j^{(b)}-\alpha^*_j)^2,
$
$$
\text{FDR}=\frac{1}{B}\sum_{b=1}^{B}\frac{\Big|\left\{j|\alpha^*_j= 0,\wh{\alpha}_j^{(b)}\neq 0\right\}\Big|}{\Big|\left\{j|\wh{\alpha}_j^{(b)}\neq 0\right\}\Big|}.
$$
We also record whether we correctly identify the existence of an unmeasured confounding effect, denoted by TIR. The definition is $$\text{TIR}=\frac{1}{B}\sum_{b=1}^{B}\left\{1\left(\UCbparahat=\bzero,\UCbpara^*=\bzero\right)+1\left(\UCbparahat\neq\bzero,\UCbpara^*\neq\bzero\right)\right\}.$$
The empirical results are based on $B=500$ replications.

The simulation results are shown in Table \ref{tabMSE} and \ref{tabTIRFDR}. 
We make the following observations. 
(i) For the Oracle estimators, the estimators that utilize RWD act better than RCT-only estimators. The proposed Oracle estimator (RL.or) has the minimal RMSE in all settings compared with the estimates from other methods.
(ii) The estimator that ignores the unmeasured confounding effect in RWD (RL.NAI) has the highest RMSE. This shows that ignorance of unmeasured confounding can lead to significant estimation error, confirming the necessity of identifying unmeasured confounding effectS in RWD.
(iii) For the methods to select tuning parameters, CV is competitive with BIC when there is no unmeasured confounding effect and better than BIC when there is unmeasured confounding effect. In the following, we just analyze the results from the CV. 
All estimators that utilize RWD have smaller RMSE than the RCT-only estimators.
Among the reported estimators, the RMSE of the proposed estimate (RL.cv) is sensibly lower than other methods. The RMSE of OA.cv is the second lowest. The RMSE of Meta.cv is slightly higher than that of OA.cv. 
(iv) The proposed estimator (RL.cv) has a slightly lower/competitive FDR than that of OA.cv, and it is sensibly lower than that of other methods. 
(v) Based on the results of TIR, it can be seen that all methods can identify the case well when there is unmeasured confounding. When no unmeasured confounding effect exists, 
the proposed and outcome-adjusted methods perform better than other methods.
(vi) Generally, the RMSEs and FDRs have no better performances when  the censoring rate increases. If there is no unmeasured confounding effect, the estimators that utilize RWD gain more efficiency than the estimators in the case where an unmeasured confounding effect exists. 

Additional simulation experiments considering a weaker signal strength of the coefficients ($Signal=1$), a more severer censoring rate (CR$=60\%$), and the log-logistic distribution of the survival time are presented in the supplementary materials in detail. The results show that the proposed method maintains its effectiveness across these  settings.
To summarize, the proposed method can  identify unmeasured confounding effects well and  gains more efficiency than the RCT-only estimators. The proposed estimator did well in cases including relatively high dimensions and severe censoring. In addition, it acts the best compared with the estimates from other reported methods.

\subsection{Variance estimation}
Simulations are implemented to evaluate the nonparametric bootstrap approach for variance estimation. 
The details of the estimation method are presented in Remark \ref{var}.
We compute the variance estimates for the proposed method using two types of data: one combining RCT with RWD (denoted as RCT+RWD), and the other using RCT data alone. 
Here we take the bootstrap sample size of 500. 
In Table \ref{tabSE} and \ref{tabSE2}, we show the average of the point estimates (Mean), standard deviations (SD), the means of the bootstrap estimated standard deviations (SE), and the 0.95 coverage proportion (CP) based on 500 replications. 

Upon examining Tables \ref{tabSE}  and \ref{tabSE2}, it is evident that the bootstrap standard deviation estimates match the standard deviations of the estimates well. Furthermore, the variance of the estimates derived from the combined RCT+RWD dataset is observed to be lower than that obtained from RCT data alone. This variance reduction, or shrinkage, is particularly pronounced for the coefficients in $\supp_1\setminus\supp_2$.

\section{Application}
Lung cancer has become the primary cause of cancer-related deaths across the globe, with increasing incidence over the last two decades (\citealp{Sung2021}).
Surgical resection, including lobectomy and sublobar resection, is commonly used for early-stage lung cancer. Lobectomy involves the complete removal of the lung lobe where the tumor is located, while sublobar resection only entails the removal of a smaller section of the complicated lobe.
In 1995, Ginsberg and Rubinstein reported a randomized trial that compared lobectomy with sublobar resection in patients with clinical T1N0 non-small-cell lung cancer (NSCLC) (\citealp{Ginsberg1995}). 
They found that compared with lobectomy, sublobar resection does not confer improved perioperative morbidity, mortality, or late postoperative pulmonary function.  
These results made lobectomy the standard of surgical treatment for patients with clinical T1N0 NSCLC. Sublobar resection for early-stage lung cancer has only been assigned for patients with poor pulmonary reserve or other major comorbidities contraindicating lobectomy. 
Over the years, however, advances in imaging and staging methods have allowed the detection of smaller and earlier tumors, leading to a renewed interest in sublobar resection for patients with clinical stage IA NSCLC who might otherwise accept a lobectomy (\citealp{Saji2022}). 

\begin{figure}[h]
	\centering		    			
	\includegraphics[width=0.75\textwidth]{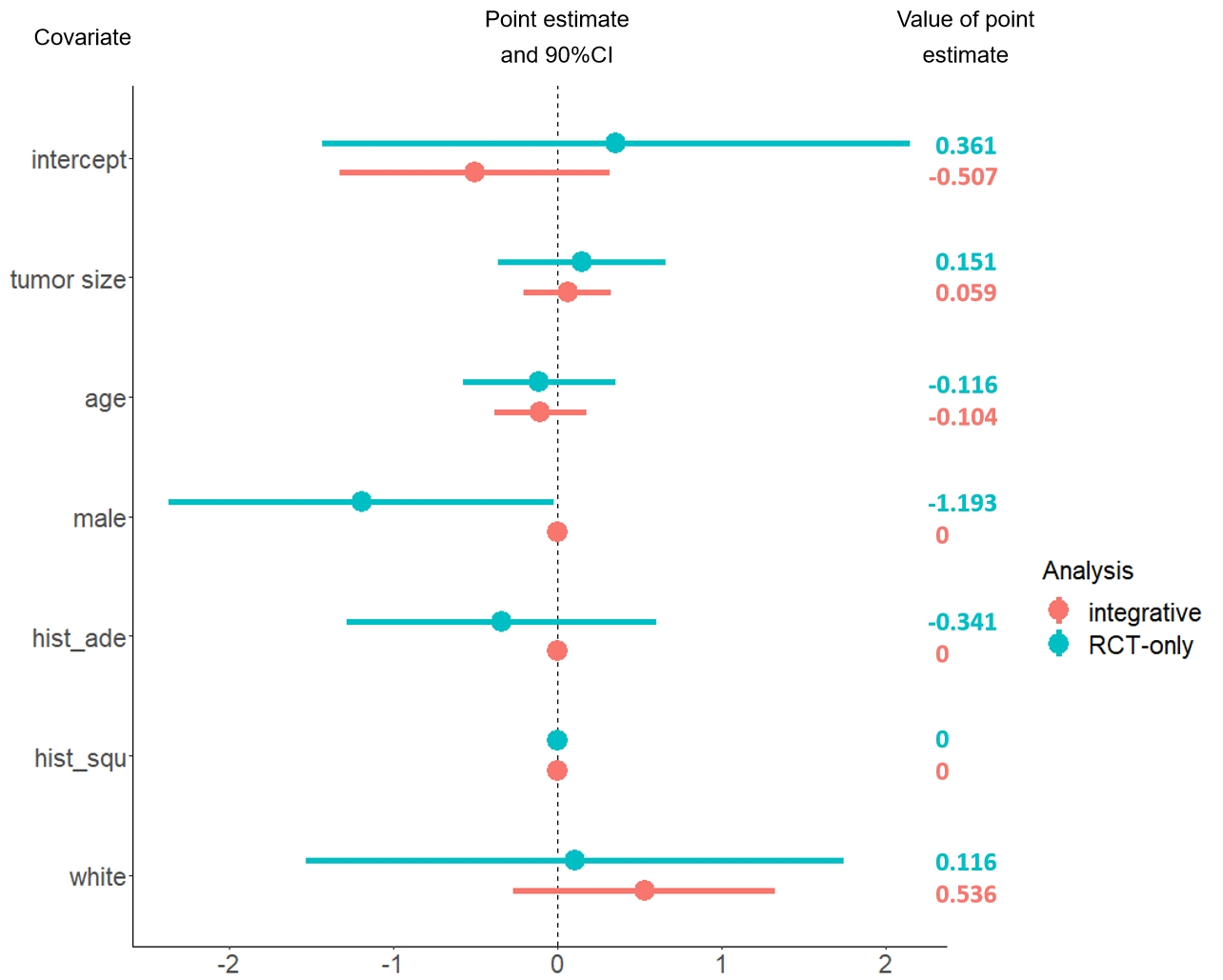}	
	\caption{The estimated covariate effects in HTE. Here $hist\_ade$ indicates a presence of histologic type - adenocarcinoma, $hist\_squ$ suggests a presence of the histologic type - squamous-cell carcinoma.}\label{appli_hte}
\end{figure}
\begin{figure}[h]
	\centering		    			
	\includegraphics[width=0.75\textwidth]{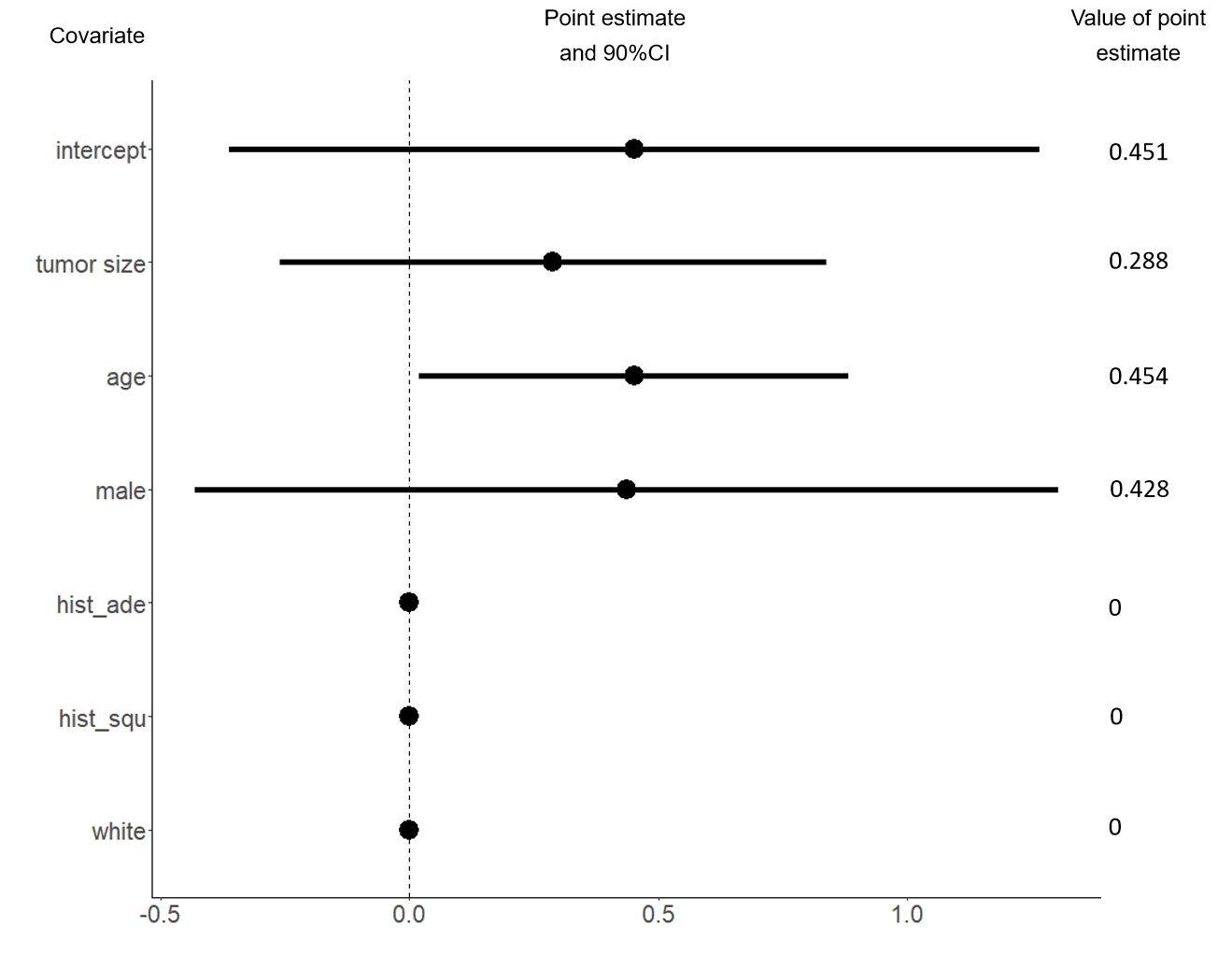}	
	\caption{The estimated covariate effects in unmeasured confounding. Here $hist\_ade$ indicates a presence of histologic type - adenocarcinoma, $hist\_squ$ suggests a presence of the histologic type - squamous-cell carcinoma.}\label{appli_uc}
\end{figure}

C140503 is a multicenter, noninferiority, phase 3 trial where NSCLC patients with tumor size $\leq$2 cm were randomly assigned to undergo sublobar resection or lobar resection after intraoperative confirmation of node-negative disease (\citealp{Altorki2023}). From June 2007 to March 2017, a total of 697 patients were assigned to undergo sublobar resection (340 patients) or lobar resection (357 patients). For disease-free survival, the right censoring rate is 59.7\% in the group with sublobar resection and 60.5\% in the group with lobar resection. It concluded that sublobar resection was non-inferior to lobar resection with respect to disease-free survival. In addition, a post hoc analysis of the heterogeneity of treatment effects for disease-free survival across patient subgroups based on Cox proportional hazards models revealed that age and tumor size intended to post a negative effect and positive effect on lobar resection, respectively. 
NCDB is a clinical oncology database maintained by the American College of Surgeons, and it accounts for 72\% of all newly diagnosed lung cancer cases in the United States. 
The NCDB analysis based on multivariate Cox proportional hazards models and propensity score-based methods reveals a significant advantage of lobectomy over limited resection, which contradicts the results of C140503. This contradictory result may be attributed to unobserved hidden confounders in the NCDB database.  It has been well-documented that surgeons and patients tend to opt for limited resection over lobectomy when the patient’s health status is poor, functional respiratory service is low, and there is a high burden of comorbidities (\citealp{Zhang2019,Lee2023}).  Unfortunately, these hidden confounders were not captured in the NCDB database, which could potentially result in biased estimates of treatment effects. 

Though NCDB provides abundant samples, it fails to give a valid result of causal effect due to unmeasured confounding. 
We intend to apply the proposed method to integrate the NCDB data to C140503. It is interesting to see whether the efficiency of the HTE estimate can be improved. 
We randomly selected a cohort of 3000 patients with stage 1A NSCLC from the NCDB database, ensuring that their tumor size was $\leq$2 cm and they met all the eligibility criteria for C140503. 
We consider the covariates that appear in both C140503 and NCDB, including race (white and other), sex (male and female), age, tumor size, histologic type (squamous-cell carcinoma, adenocarcinoma and other).  
The estimated covariate effects in HTE and unmeasured confounding are presented in Figure \ref{appli_hte} and \ref{appli_uc} respectively.
In Figure \ref{appli_uc}, the result shows that the effect of unmeasured confounding exists, which is consistent with the previous findings.  It also reveals that the hidden confounding is significantly related to the patient's age under 90\% confidence level. 
In Figure \ref{appli_hte}, it can be observed that compared with the C140503-only method, the proposed integrative estimator yields shorter confidence intervals. 
In particular, the estimated effects of sex ($\alpha_{sex}$) and presence of histologic type adenocarcinoma ($\alpha_{ade}$)  are shrunk to zero when integrating NCDB. Since the C140503-only estimates $\wh{\alpha}_{sex}^{rct}$, $\wh{\alpha}_{ade}^{rct}$ show that the upper tail of the 90\% confidence interval of $\wh{\alpha}_{sex}^{rct}$ is closed to zero and $\wh{\alpha}_{ade}^{rct}$ is not sigificant, it is reasonable to see the shrinkage when sythesizing NCDB. 
These results indicate that integrating the NCDB data to C140503 does improve the efficiency, which convincingly demonstrates the practical effectiveness of the proposed method.

\section{Conclusion}
In this paper, we have developed an integrative method to give an improved estimate of HTE by synthesizing the evidence from RCTs and RWD, particularly in situations where the outcome of interest is subject to censoring and the number of covariates is diverging. It can be seen that the situations we consider are more complex and realistic, bringing more challenges. The proposed method can deal with cases where unmeasured confounding is present in RWD. It can identify whether the unmeasured confounding effect exists in a fully data-driven manner, contributing to more efficient estimates and a deeper understanding of the data generation mechanism. 
We have rigorously established the theoretical properties, showing that the proposed integrative method yields a more efficient HTE estimate, at least as good as those based on only the RCTs data. 
The proposed method is practically applicable. Based on the evidence from C140503, the randomized controlled trial, and NCDB database, the real-world data that might be subject to hidden confounding, we have applied the proposed method to improve the estimate of the HTE on survival for patients with clinical T1N0 NSCLC undergoing lobar resection. The results reported that integrating NCDB data into C140503 enhanced,  convincingly indicating the practicality of the proposed method. 

In this project, we focus on developing data integration methods utilizing Stute weights, given their widespread use and suitability under the censoring assumptions. 
	However, we acknowledge the potential benefits of exploring more general doubly robust weighting approaches. Future work could extend our methods to incorporate IPCW and doubly robust techniques, potentially building on the frameworks established by \citet{Lee2022} and \citet{Lee2024}, initially designed for trial generalization.
Moreover, in the context of integrating RWD into RCTs, there are still many problems to be solved. For example, it is common to see that RCTs and RWD have different covariates. Merely taking into account the shared covariates may incur other problems. For instance, some critical covariates to describe heterogeneity in treatment may be excluded. Thus, it is important to develop an integrative approach that can deal with non-uniform covariates in RCTs and RWD.  
Moreover, RCTs with time-varying treatments are common. Integrative analysis of the continuous-time structural failure time model (\citealp{YangAFT2020}) combining the complementary features of RCT and RWD will be an important topic for future research.


\clearpage

\begin{table}[h]\footnotesize
	\setstretch{1.2}
	\setlength{\tabcolsep}{2mm}
	\begin{threeparttable}
		\caption{The RMSE ($\times10^{2}$) of the HTE estimation when $Signal=2$ over 500 experiment replicates}\label{tabMSE}
		\begin{tabular}{lcccccccc}
			\toprule
			\multicolumn{1}{c}{\multirow{3}{*}{Methods}} & \multicolumn{4}{c}{with unmeasured confounding}                                      & \multicolumn{4}{c}{no unmeasured confounding}                                         \\
			\cmidrule(r){2-5}
			\cmidrule(r){6-9}
			\multicolumn{1}{c}{}                         & \multicolumn{2}{c}{p=20}                  & \multicolumn{2}{c}{p=50}                  & \multicolumn{2}{c}{p=20}                  & \multicolumn{2}{c}{p=50}                  \\
			\cmidrule(r){2-3}
			\cmidrule(r){4-5}
			\cmidrule(r){6-7}
			\cmidrule(r){8-9}
			\multicolumn{1}{c}{}                         & CR=20\%             & CR=40\%            & CR=20\%             & CR=40\%             & CR=20\%             & CR=40\%             & CR=20\%             & CR=40\%             \\
			\midrule
			\textbf{RL.or}      & 8.24                & 10.08                & 5.29                & 6.64                & 5.42                & 6.71                & 3.55                & 4.22                \\
			\textbf{RL.RCTor}   & 15.16                & 18.15                & 9.22                 & 11.07                & 15.20                & 17.94                & 8.96                & 11.09               \\
			\textbf{RL.NAIor}   & 71.00                & 71.10                & 44.83                & 44.98                &               &             &               &             \\
			\textbf{RL.cv}      & {\bf8.40} & {\bf10.35} & \uline {{\bf5.31}} & \uline{{\bf6.71}} & {\bf5.64} & {\bf7.15} & {\bf3.59} & {\bf4.35} \\
			\textbf{RL.bic}     & \uline {8.26}                & \uline{10.13}   & 5.32                & 7.63                & \uline{5.48}                & \uline{6.81 }               & \uline{3.56}               & \uline{4.19}               \\
			\textbf{RL.RCT}     & 15.67                & 19.05                & 9.29                & 11.78               & 15.71                & 18.95                & 9.22                & 11.82               \\
			\textbf{RL.NAI}     & 71.67                & 72.28                & 45.88                & 46.67                &                     &                     &                     &                     \\
			\midrule
			\textbf{OA.or}      & 11.30                & 12.97                & 7.10                & 8.54                 & 8.87                 & 10.72                & 5.71                & 7.04                \\
			\textbf{OA.RCTor}   & 22.23                & 25.97                & 15.82               & 18.49                & 22.32                & 25.95                & 15.75                & 18.57                \\
			\textbf{OA.cv}      & {\bf11.34 }      & {\bf13.10 }       & {\bf7.13 }        & {\bf8.60}       & {\bf8.90 }     & {\bf10.81}    & {\bf5.73}    & {\bf7.05}      \\
			\textbf{OA.bic}     &  11.33                & 13.03               & 7.13               & 9.12                 & 8.89                 & 10.75                & 5.72                & 7.04               \\
			\textbf{OA.RCT}     & 22.36                & 26.20                & 15.89                & 18.65                & 22.44                & 26.20                & 15.84               & 18.73                \\
			\midrule
			\textbf{GM0.or}     & 13.73                & 15.87                & 8.81                 & 10.33                & 10.47                & 12.43                & 6.76                & 8.06                \\
			\textbf{GM0.RCTor}  & 24.92                & 29.09                & 17.40                & 20.61                & 24.92                & 29.09                & 17.40                & 20.42                \\
			\textbf{GM0.cv}     &  {\bf14.26}        & {\bf16.69}       & {\bf9.03}       & {\bf10.81   }    & {\bf11.16 }     & {\bf13.13}        & {\bf6.98}      & {\bf8.45}       \\
			\textbf{GM0.biv}    &13.78               & 16.20                & 8.97                 & 11.41                & 10.74                & 12.54                & 6.85                & 8.10                \\
			\textbf{GM0.RCT}    & 24.81                & 29.10                & 17.35                & 21.02                & 24.81                & 29.10                & 17.32                & 20.85                \\
			\midrule
			\textbf{GM1.or}     & 11.76                & 13.79                & 7.54                & 9.23                 & 9.52                 & 11.35                & 6.13                & 7.40                \\
			\textbf{GM1.RCTor}  & 22.67                & 26.66               & 15.84                & 18.36                & 22.62                & 26.62                & 15.93                & 18.36                \\
			\textbf{GM1.cv}     & {\bf12.35}       & {\bf14.68}       &{\bf7.81}      &{\bf9.86}        &{\bf9.99 }      & {\bf11.98}       &{\bf6.34 }      &{\bf7.77}       \\
			\textbf{GM1.bic}    & 11.93                & 14.02                & 7.62                & 10.16                & 9.70                 &11.50               & 6.22                & 7.46                \\
			\textbf{GM1.RCT}    & 22.57                & 26.64                & 15.77                & 18.51                & 22.51                & 26.61                & 15.88                & 18.48                \\
			\midrule
			\textbf{Meta.or}    & 11.14                & 12.86                & 7.16                 & 8.49                 & 9.32                 & 11.12                & 6.02                & 7.24                \\
			\textbf{Meta.RCTor} & 22.34                & 25.93                & 15.68                & 18.19                & 22.31                & 25.93                & 15.67                & 18.12                \\
			\textbf{Meta.cv}    & {\bf11.48 }      & {\bf13.39}       & {\bf7.31 }     & {\bf8.86 }       &  {\bf9.70}        & {\bf11.45}      & {\bf6.18 }      & {\bf7.45 }  \\
			\textbf{Meta.bic}   & 11.21                & 13.04                & 7.25                 & 9.20               &9.51                 & 11.20               & 6.11                & 7.27                \\
			\textbf{Meta.RCT}   & 22.17                & 25.77                & 15.59                & 18.35                & 22.13                & 25.77                & 15.58                & 18.28                \\
			\midrule
			\textbf{GM01.or}    & 13.77                & 17.05                & 8.95                 & 11.30                & 11.50                & 14.86                & 7.49                 & 9.77                 \\
			\textbf{GM01.RCTor} & 28.50                & 36.42                & 19.88                & 25.77                & 28.41                & 36.36                & 19.93                & 25.56                \\
			\textbf{GM01.cv}    & {\bf14.63}        & {\bf18.69 }       & {\bf9.66}        & {\bf13.40}      & {\bf12.49}        & {\bf16.61}       & {\bf8.06 }       & {\bf11.80}       \\
			\textbf{GM01.bic}   & 13.94               & 17.48               & 9.03                 & 11.60               & 11.83               & 15.50                & 7.54                 & 10.09                \\
			\textbf{GM01.RCT}   & 28.65                & 36.67                & 19.95                & 26.03                & 28.56                & 36.60                & 19.99                & 25.77                           \\
			\bottomrule  
		\end{tabular}
		\footnotesize{
			Some results are marked in bold to make it clear for readers to make comparison between different methods. The results behave the best except for the oracle estimates (i.e., smallest RMSE) are marked with underlines.\\
				In the table, CR represents cencoring rate. 
				Among these methods, those with names starting with ``RL" indicate the proposed model. RL.cv and RL.bic represent the proposed estimates under CV and BIC criterion respectively. RL.RCT represents the estimate merely based on RCT data. RL.NAI is the naive estimate which completely ignores unmeasured confounding effect. 
				RL.or, RL.RCTor, and RL.NAIor are the oracle estimates of the integrative analysis, RCT-only analysis, and naive analysis respectively. 
				Other methods with names starting with ``OA'', ``GM0'', ``GM1'', ``Meta'' and ``GM01'' are introduced in detail in Section 4.
		}
	\end{threeparttable}
\end{table}

\begin{table}[h]\footnotesize
	\setstretch{1.2}
	\setlength{\tabcolsep}{1.5mm}
	\begin{threeparttable}
		\caption{The averaged TIR($/\%$) and FDR($/\%$) when Signal=2 over 500 experiment replicates.}\label{tabTIRFDR}
		\begin{tabular}{clcccccccc}
			\toprule
			\multirow{3}{*}{Index} & \multicolumn{1}{c}{\multirow{3}{*}{Methods}} & \multicolumn{4}{c}{with unmeasured confounding}     & \multicolumn{4}{c}{no unmeasured confounding}       \\
			\cmidrule(r){3-6}
			\cmidrule(r){7-10}
			& \multicolumn{1}{c}{}                         & \multicolumn{2}{c}{p=20} & \multicolumn{2}{c}{p=50} & \multicolumn{2}{c}{p=20} & \multicolumn{2}{c}{p=50} \\
			\cmidrule(r){3-4}
			\cmidrule(r){5-6}
			\cmidrule(r){7-8}
			\cmidrule(r){9-10}
			& \multicolumn{1}{c}{}                         & CR=20\%     & CR=40\%    & CR=20\%     & CR=40\%    & CR=20\%     & CR=40\%    & CR=20\%     & CR=40\%    \\
			\midrule
			\multirow{12}{*}{TIR}  
			& \textbf{RL.cv}    & 100 & 100 & 100 & 100 & 97.8 & 97.3 & 98.5 & 98.1 \\
			&\textbf{RL.bic}   & 100 & 100 & 100 & 100 & 97.8 & 98.6 & 99.2 & 98.8 \\
			\cmidrule{2-10}
			&\textbf{OA.cv}    & 100 & 100 & 100 & 100 & 98.8 & 97.8 & 99.2 & 99.0 \\
			&\textbf{OA.bic}   & 100 & 100 & 100 & 100 & 99.4 & 99.4 & 99.2 & 99.1 \\
			\cmidrule{2-10}
			&\textbf{GM0.cv}   & 100 & 100 & 100 & 100 & 85.0 & 82.0 & 90.8 & 87.2 \\
			&\textbf{GM0.biv}  & 100 & 100 & 100 & 100 & 95.4 & 95.4 & 97.8 & 97.6 \\
			\cmidrule{2-10}
			&\textbf{GM1.cv}   & 100 & 100 & 100 & 100 & 87.0 & 84.6 & 92.4 & 86.8 \\
			&\textbf{GM1.bic}  & 100 & 100 & 100 & 100 & 96.6 & 94.2 & 98.4 & 95.2 \\
			\cmidrule{2-10}
			&\textbf{Meta.cv}  & 100 & 100 & 100 & 100 & 74.2 & 70.2 & 85.0 & 75.4 \\
			&\textbf{Meta.bic} & 100 & 100 & 100 & 100 & 92.0 & 89.6 & 96.2 & 92.8 \\
			\cmidrule{2-10}
			&\textbf{GM01.cv}  & 100 & 100 & 100 & 100 & 60.2 & 26.8 & 61.6 & 13.4 \\
			&\textbf{GM01.bic} & 100 & 100 & 100 & 100 & 88.4 & 67.8 & 92.6 & 64.2  \\
			\midrule
			\multirow{19}{*}{FDR} & \textbf{RL.cv}    & 0.28 & 0.67 & 0.07 & 0.71 & 0.32 & 0.93 & 0.13 & 0.37 \\
			& \textbf{RL.bic}   & 0.07 & 0.20 & 0.04 & 0.51 & 0.07 & 0.20 & 0.02 & 0.09 \\
			& \textbf{RL.RCT}   & 3.51 & 5.00 & 3.48 & 6.16 & 3.39 & 5.07 & 3.63 & 5.81 \\
			& \textbf{RL.NAI}   & 1.31 & 2.61 & 0.87 & 2.28 &      &      &      &      \\
			\cmidrule{2-10}
			& \textbf{OA.cv}    & 0.04 & 0.11 & 0.04 & 0.18 & 0.09 & 0.37 & 0.04 & 0.07 \\
			& \textbf{OA.bic}   & 0.04 & 0.04 & 0.02 & 0.11 & 0.02 & 0.09 & 0.00 & 0.00 \\
			& \textbf{OA.RCT}   & 0.90 & 1.55 & 0.56 & 1.59 & 0.80 & 1.71 & 0.74 & 1.71 \\
			\cmidrule{2-10}
			& \textbf{GM0.cv}   & 1.37 & 2.71 & 1.63 & 3.85 & 0.90 & 1.91 & 1.15 & 3.00 \\
			& \textbf{GM0.biv}  & 0.18 & 0.77 & 0.40 & 1.34 & 0.24 & 0.34 & 0.16 & 0.24 \\
			& \textbf{GM0.RCT}  & 1.51 & 2.24 & 0.86 & 1.81 & 1.51 & 2.24 & 1.03 & 1.65 \\
			\cmidrule{2-10}
			& \textbf{GM1.cv}   & 1.14 & 2.26 & 0.92 & 3.35 & 0.68 & 1.49 & 0.84 & 2.51 \\
			& \textbf{GM1.bic}  & 0.13 & 0.59 & 0.07 & 1.18 & 0.16 & 0.18 & 0.11 & 0.24 \\
			& \textbf{GM1.RCT}  & 1.21 & 2.55 & 0.46 & 1.85 & 1.29 & 2.66 & 0.68 & 1.44 \\
			\cmidrule{2-10}
			& \textbf{Meta.cv}  & 2.45 & 4.71 & 2.50 & 6.80 & 1.54 & 3.24 & 1.98 & 5.23 \\
			& \textbf{Meta.bic} & 0.31 & 1.35 & 0.46 & 2.43 & 0.37 & 0.52 & 0.27 & 0.46 \\
			& \textbf{Meta.RCT} & 2.66 & 4.63 & 1.31 & 3.58 & 2.76 & 4.74 & 1.71 & 3.05 \\
			\cmidrule{2-10}
			& \textbf{GM01.cv}  & 1.69 & 7.05 & 3.99 & 19.3 & 1.38 & 5.50 & 2.51 & 16.6 \\
			& \textbf{GM01.bic} & 0.16 & 1.02 & 0.28 & 2.24 & 0.18 & 1.08 & 0.13 & 2.24 \\
			& \textbf{GM01.RCT} & 1.18 & 2.30 & 0.37 & 1.81 & 1.12 & 2.09 & 0.30 & 1.60 \\
			\bottomrule
		\end{tabular}
		\footnotesize{
				In the table, CR represents cencoring rate. TIR is the rate of correctly identifying the real case where unmeasured confounding effect exists or not. FDR is the false discovery rate of the HTE estimates. 
				Among these methods, those with names starting with ``RL" indicate the proposed model. RL.cv and RL.bic represent the proposed estimates under CV and BIC criterion respectively. RL.RCT represents the estimate merely based on RCT data. RL.NAI is the naive estimate which completely ignores unmeasured confounding effect. 
				Other methods with names starting with ``OA'', ``GM0'', ``GM1'', ``Meta'' and ``GM01'' are introduced in detail in Section 4. }
	\end{threeparttable}
\end{table}

\begin{table}[h]\footnotesize
	\setstretch{1.2}
	\setlength{\tabcolsep}{2mm}
	\begin{threeparttable}
		\caption{The inference results of the proposed HTE estimate over 500 experiment replicates when $p=20$}\label{tabSE}
		\begin{tabular}{cccllllllll}
			\toprule
			Case               & Dataset                  & Index    & \multicolumn{1}{c}{$\alpha^*_1=-2$} & \multicolumn{1}{c}{$\alpha^*_2=-2$} & \multicolumn{1}{c}{$\alpha^*_3=-2$} & \multicolumn{1}{c}{$\alpha^*_4=-2$} & \multicolumn{1}{c}{$\alpha^*_5=2$} & \multicolumn{1}{c}{$\alpha^*_6=2$} & \multicolumn{1}{c}{$\alpha^*_7=2$} & \multicolumn{1}{c}{$\alpha^*_8=2$} \\
			\midrule
			\multirow{8}{*}{1} & \multirow{4}{*}{RCT}     & Bias     & 0.090                     & 0.098                     & 0.098                     & 0.083                     & -0.079                   & -0.102                   & -0.095                   & -0.084                   \\
			&                          & SD       & 0.167                     & 0.185                     & 0.193                     & 0.193                     & 0.192                    & 0.190                    & 0.182                    & 0.177                    \\
			&                          & SE       & \textbf{0.188}            & \textbf{0.196}            & \textbf{0.196}            & \textbf{0.203}            & \textbf{0.197}           & \textbf{0.197}           & \textbf{0.196}           & \textbf{0.192}           \\
			&                          & CP(95\%) & 0.939                     & 0.921                     & 0.926                     & 0.936                     & 0.934                    & 0.933                    & 0.933                    & 0.939                    \\
			\cmidrule{2-11}
			& \multirow{4}{*}{RCT+RWD} & Bias     & 0.008                     & 0.006                     & 0.006                     & 0.035                     & 0.010                    & 0.005                    & 0.005                    & 0.013                    \\
			&                          & SD       & 0.080                     & 0.085                     & 0.083                     & 0.094                     & 0.086                    & 0.087                    & 0.083                    & 0.079                    \\
			&                          & SE       & \textbf{0.081}            & \textbf{0.084}            & \textbf{0.083}            & \textbf{0.090}            & \textbf{0.088}           & \textbf{0.084}           & \textbf{0.084}           & \textbf{0.081}           \\
			&                          & CP(95\%) & 0.946                     & 0.947                     & 0.923                     & 0.910                     & 0.931                    & 0.931                    & 0.944                    & 0.944                    \\
			\midrule
			\multirow{8}{*}{2} & \multirow{4}{*}{RCT}     & Bias     & 0.079                     & 0.076                     & 0.088                     & 0.073                     & -0.083                   & -0.079                   & -0.094                   & -0.079                   \\
			&                          & SD       & 0.208                     & 0.223                     & 0.234                     & 0.233                     & 0.240                    & 0.214                    & 0.215                    & 0.221                    \\
			&                          & SE       & \textbf{0.230}            & \textbf{0.239}            & \textbf{0.239}            & \textbf{0.247}            & \textbf{0.239}           & \textbf{0.241}           & \textbf{0.239}           & \textbf{0.232}           \\
			&                          & CP(95\%) & 0.959                     & 0.959                     & 0.928                     & 0.949                     & 0.925                    & 0.959                    & 0.955                    & 0.949                    \\
			\cmidrule{2-11}
			& \multirow{4}{*}{RCT+RWD} & Bias     & 0.009                     & 0.006                     & 0.009                     & 0.035                     & 0.010                    & 0.007                    & 0.004                    & 0.012                    \\
			&                          & SD       & 0.094                     & 0.100                     & 0.101                     & 0.114                     & 0.100                    & 0.101                    & 0.103                    & 0.091                    \\
			&                          & SE       & \textbf{0.097}            & \textbf{0.099}            & \textbf{0.101}            & \textbf{0.110}            & \textbf{0.111}           & \textbf{0.100}           & \textbf{0.100}           & \textbf{0.099}           \\
			&                          & CP(95\%) & 0.955                     & 0.951                     & 0.940                     & 0.904                     & 0.947                    & 0.932                    & 0.957                    & 0.947                    \\
			\midrule
			\multirow{8}{*}{3} & \multirow{4}{*}{RCT}     & Bias     & 0.102                     & 0.092                     & 0.099                     & 0.078                     & -0.076                   & -0.098                   & -0.086                   & -0.098                   \\
			&                          & SD       & 0.172                     & 0.184                     & 0.194                     & 0.197                     & 0.187                    & 0.190                    & 0.185                    & 0.182                    \\
			&                          & SE       & \textbf{0.187}            & \textbf{0.196}            & \textbf{0.196}            & \textbf{0.202}            & \textbf{0.197}           & \textbf{0.198}           & \textbf{0.198}           & \textbf{0.191}           \\
			&                          & CP(95\%) & 0.933                     & 0.931                     & 0.927                     & 0.931                     & 0.944                    & 0.929                    & 0.944                    & 0.929                    \\
			\cmidrule{2-11}
			& \multirow{4}{*}{RCT+RWD} & Bias     & 0.040                     & 0.021                     & 0.030                     & -0.015                    & 0.004                    & 0.006                    & 0.007                    & 0.011                    \\
			&                          & SD       & 0.153                     & 0.160                     & 0.160                     & 0.164                     & 0.081                    & 0.082                    & 0.088                    & 0.080                    \\
			&                          & SE       & \textbf{0.169}            & \textbf{0.182}            & \textbf{0.166}            & \textbf{0.163}            & \textbf{0.090}           & \textbf{0.093}           & \textbf{0.094}           & \textbf{0.089}           \\
			&                          & CP(95\%) & 0.958                     & 0.960                     & 0.949                     & 0.947                     & 0.967                    & 0.960                    & 0.956                    & 0.960                    \\
			\midrule
			\multirow{8}{*}{4} & \multirow{4}{*}{RCT}     & Bias     & 0.091                     & 0.081                     & 0.087                     & 0.069                     & -0.084                   & -0.080                   & -0.093                   & -0.078                   \\
			&                          & SD       & 0.205                     & 0.221                     & 0.231                     & 0.228                     & 0.228                    & 0.215                    & 0.208                    & 0.224                    \\
			&                          & SE       & \textbf{0.230}            & \textbf{0.241}            & \textbf{0.238}            & \textbf{0.245}            & \textbf{0.239}           & \textbf{0.239}           & \textbf{0.238}           & \textbf{0.231}           \\
			&                          & CP(95\%) & 0.959                     & 0.953                     & 0.931                     & 0.957                     & 0.936                    & 0.949                    & 0.949                    & 0.938                    \\
			\cmidrule{2-11}
			& \multirow{4}{*}{RCT+RWD} & Bias     & 0.041                     & 0.032                     & 0.030                     & -0.027                    & 0.011                    & 0.012                    & 0.011                    & 0.011                    \\
			&                          & SD       & 0.188                     & 0.185                     & 0.188                     & 0.187                     & 0.103                    & 0.104                    & 0.104                    & 0.100                    \\
			&                          & SE       & \textbf{0.216}            & \textbf{0.230}            & \textbf{0.204}            & \textbf{0.202}            & \textbf{0.122}           & \textbf{0.123}           & \textbf{0.122}           & \textbf{0.118}           \\
			&                          & CP(95\%) & 0.966                     & 0.972                     & 0.968                     & 0.968                     & 0.966                    & 0.976                    & 0.970                    & 0.968          \\
			\bottomrule         
		\end{tabular}
		\footnotesize{Case 1-4 represent (nuc, 20\%CR), (nuc, 40\%CR), (uc, 20\%CR), (uc, 40\%CR), where nuc means there is no unmeasured confounding while uc means there is unmeasured confounding.
			The SEs are marked in bold to make it clear for readers to make comparison between RCT and RCT+RWD.}
	\end{threeparttable}
\end{table}

\begin{table}[h]\footnotesize
	\setstretch{1.2}
	\setlength{\tabcolsep}{2mm}
	\begin{threeparttable}
		\caption{The inference results of the proposed HTE estimate over 500 experiment replicates when $p=50$}\label{tabSE2}
		\begin{tabular}{cccllllllll}
			\toprule
			Case               & Dataset                  & Index    & \multicolumn{1}{c}{$\alpha^*_1=-2$} & \multicolumn{1}{c}{$\alpha^*_2=-2$} & \multicolumn{1}{c}{$\alpha^*_3=-2$} & \multicolumn{1}{c}{$\alpha^*_4=-2$} & \multicolumn{1}{c}{$\alpha^*_5=2$} & \multicolumn{1}{c}{$\alpha^*_6=2$} & \multicolumn{1}{c}{$\alpha^*_7=2$} & \multicolumn{1}{c}{$\alpha^*_8=2$} \\
			\midrule
			\multirow{8}{*}{1} & \multirow{4}{*}{RCT}     & Bias    & 0.076          & 0.073          & 0.071          & 0.067          & -0.082         & -0.056         & -0.084         & -0.083         \\
			&                          & SD      & 0.170          & 0.176          & 0.180          & 0.189          & 0.189          & 0.180          & 0.183          & 0.177          \\
			&                          & SE      & \textbf{0.180} & \textbf{0.189} & \textbf{0.189} & \textbf{0.196} & \textbf{0.188} & \textbf{0.189} & \textbf{0.188} & \textbf{0.182} \\
			&                          & CP(95\%) & 0.934          & 0.948          & 0.950          & 0.946          & 0.924          & 0.962          & 0.928          & 0.940          \\
			\cmidrule{2-11}
			& \multirow{4}{*}{RCT+RWD} & Bias    & 0.000          & 0.015          & 0.008          & 0.045          & 0.001          & 0.003          & 0.003          & 0.006          \\
			&                          & SD      & 0.084          & 0.085          & 0.089          & 0.086          & 0.085          & 0.089          & 0.089          & 0.083          \\
			&                          & SE      & \textbf{0.079} & \textbf{0.082} & \textbf{0.082} & \textbf{0.086} & \textbf{0.085} & \textbf{0.082} & \textbf{0.082} & \textbf{0.082} \\
			&                          & CP(95\%) & 0.942          & 0.930          & 0.930          & 0.903          & 0.938          & 0.928          & 0.930          & 0.946          \\
			\midrule
			\multirow{8}{*}{2} & \multirow{4}{*}{RCT}     & Bias    & 0.052          & 0.076          & 0.079          & 0.068          & -0.068         & -0.058         & -0.059         & -0.072         \\
			&                          & SD      & 0.201          & 0.220          & 0.217          & 0.224          & 0.218          & 0.216          & 0.229          & 0.221          \\
			&                          & SE      & \textbf{0.222} & \textbf{0.232} & \textbf{0.232} & \textbf{0.240} & \textbf{0.232} & \textbf{0.232} & \textbf{0.232} & \textbf{0.224} \\
			&                          & CP(95\%) & 0.948          & 0.952          & 0.954          & 0.958          & 0.947          & 0.958          & 0.948          & 0.947          \\
			\cmidrule{2-11}
			& \multirow{4}{*}{RCT+RWD} & Bias    & 0.000          & 0.011          & 0.011          & 0.042          & 0.006          & 0.006          & 0.008          & 0.015          \\
			&                          & SD      & 0.093          & 0.097          & 0.101          & 0.103          & 0.098          & 0.099          & 0.104          & 0.096          \\
			&                          & SE      & \textbf{0.096} & \textbf{0.099} & \textbf{0.097} & \textbf{0.107} & \textbf{0.107} & \textbf{0.100} & \textbf{0.099} & \textbf{0.099} \\
			&                          & CP(95\%) & 0.950          & 0.948          & 0.937          & 0.906          & 0.941          & 0.941          & 0.924          & 0.933          \\
			\midrule
			\multirow{8}{*}{3} & \multirow{4}{*}{RCT}     & Bias    & 0.064          & 0.092          & 0.069          & 0.085          & -0.072         & -0.070         & -0.084         & -0.077         \\
			&                          & SD      & 0.166          & 0.188          & 0.181          & 0.189          & 0.187          & 0.184          & 0.178          & 0.177          \\
			&                          & SE      & \textbf{0.180} & \textbf{0.188} & \textbf{0.187} & \textbf{0.195} & \textbf{0.188} & \textbf{0.189} & \textbf{0.187} & \textbf{0.182} \\
			&                          & CP(95\%) & 0.943          & 0.927          & 0.937          & 0.933          & 0.945          & 0.927          & 0.931          & 0.947          \\
			\cmidrule{2-11}
			& \multirow{4}{*}{RCT+RWD} & Bias    & 0.037          & 0.041          & 0.027          & -0.014         & 0.005          & 0.007          & 0.004          & 0.008          \\
			&                          & SD      & 0.171          & 0.163          & 0.161          & 0.157          & 0.085          & 0.088          & 0.094          & 0.085          \\
			&                          & SE      & \textbf{0.177} & \textbf{0.198} & \textbf{0.170} & \textbf{0.167} & \textbf{0.090} & \textbf{0.092} & \textbf{0.091} & \textbf{0.088} \\
			&                          & CP(95\%) & 0.965          & 0.961          & 0.963          & 0.963          & 0.955          & 0.965          & 0.937          & 0.935          \\
			\midrule
			\multirow{8}{*}{4} & \multirow{4}{*}{RCT}     & Bias    & 0.067          & 0.060          & 0.070          & 0.069          & -0.066         & -0.050         & -0.077         & -0.080         \\
			&                          & SD      & 0.197          & 0.231          & 0.202          & 0.226          & 0.212          & 0.208          & 0.228          & 0.224          \\
			&                          & SE      & \textbf{0.223} & \textbf{0.235} & \textbf{0.233} & \textbf{0.240} & \textbf{0.232} & \textbf{0.232} & \textbf{0.231} & \textbf{0.223} \\
			&                          & CP(95\%) & 0.967          & 0.947          & 0.971          & 0.967          & 0.965          & 0.963          & 0.941          & 0.940          \\
			\cmidrule{2-11}
			& \multirow{4}{*}{RCT+RWD} & Bias    & 0.035          & 0.032          & 0.030          & -0.017         & 0.009          & 0.009          & 0.005          & 0.011          \\
			&                          & SD      & 0.182          & 0.209          & 0.204          & 0.199          & 0.104          & 0.109          & 0.108          & 0.103          \\
			&                          & SE      & \textbf{0.237} & \textbf{0.271} & \textbf{0.217} & \textbf{0.206} & \textbf{0.115} & \textbf{0.116} & \textbf{0.117} & \textbf{0.116} \\
			&                          & CP(95\%) & 0.974          & 0.976          & 0.943          & 0.958          & 0.960          & 0.943          & 0.949          & 0.958          \\
			\bottomrule         
		\end{tabular}
		\footnotesize{Case 1-4 represent (nuc, 20\%CR), (nuc, 40\%CR), (uc, 20\%CR), (uc, 40\%CR), where nuc means there is no unmeasured confounding while uc means there is unmeasured confounding. The SEs are marked in bold to make it clear for readers to make comparison between RCT and RCT+RWD.}
	\end{threeparttable}
\end{table}

\clearpage
\section*{Appendix A: technical details for Section 3}
Firstly, we rewrite the regression model with more concise notations. Let $\bY_i=\log(\failureT_i)-\MeanOutcome(\XS_i)$,  $\bYhat_i=\log(\failureT_i)-\MeanOutcomehat(\XS_i)$, $\bY=\left(\bY_i,i=1,2,...,\numAll\right)$, and $\bYhat=\left(\bYhat_i,i=1,2,...,\numAll\right)$.
Denote that
$$
\Lmb=\diag\left\{\Treat_i-\propen(\XS_i),i=1,2,...,\numAll\right\},\quad\Lmbhat=\diag\left\{\Treat_i-\propenhat(\XS_i),i=1,2,...,\numAll\right\}.
$$
Let $\rho(\Allbpara)=\sum_{j=1}^{\alldim}\MCP(\vert\Allpara_j\vert;\tuneHTE)+\MCP(\Vert\Allbpara_{\setUC}\Vert;\tuneUC).$ Then the loss function (\ref{loss}) can be written into
$$
\loss_{\tuneHTE,\tuneUC}(\Allbpara,\nuisanhat|\alldata)=\Big\Vert\bW^{1/2}\left(\bYhat-\Lmbhat\AllX\Allbpara\right)\Big\Vert_2^2+\rho(\Allbpara).
$$

\subsection*{Proof of Theorem \ref{thm1}}
The proof follows the lines of \citet{Fan2004}. We aim to show that for any given $\epsilon$, there is a large enough constant $M$ such that, for large $\numAll$, we have
$$
\Pr\left\{
\min_{||\dAllbpara||_2=M}\loss_{\tuneHTE,\tuneUC}(\Allbpara^*+\rate\dAllbpara,\nuisanhat|\alldata)>\loss_{\tuneHTE,\tuneUC}(\Allbpara^*,\nuisanhat|\alldata)
\right\}\geq 1-\epsilon,
$$
where $\dAllbpara\in \R^{2\alldim}$ is an increment vector. 
This implies that with probability converging to 1, there is a local minimum $\Allbparahat$ such that $\Vert\Allbparahat-\Allbpara^*\Vert_2=\OP(1)\rate$.
It can be seen that
\begin{equation*}
	\begin{split}
		&\loss_{\tuneHTE,\tuneUC}(\Allbpara^*+\rate\dAllbpara,\nuisanhat|\alldata)-\loss_{\tuneHTE,\tuneUC}(\Allbpara^*,\nuisanhat|\alldata)\\
		&=\Big\Vert\bW^{1/2}\Lmbhat\AllX\dAllbpara\Big\Vert_2^2\rate^2+2\dAllbpara\AllX^\rmT\Lmbhat\bW\left(\Lmbhat\AllX\Allbpara^*-\bYhat\right)\rate+\left\{\rho(\Allbpara^*+\rate\dAllbpara)-\rho(\Allbpara^*)\right\}\\
		&=\partI+\partII+\partIII.
	\end{split}
\end{equation*}

We first deal with $\partI$. It can be deduced that
\begin{equation*}
	\begin{split}
		\partI&=\rate^2\Big\Vert\bW^{1/2}\Lmb\AllX\dAllbpara\Big\Vert_2^2+\rate^2\Big\Vert\bW^{1/2}(\Lmb-\Lmbhat)\AllX\dAllbpara\Big\Vert_2^2+2\rate^2\dAllbpara^\rmT\AllX^\rmT\Lmb\bW(\Lmbhat-\Lmb)\AllX\dAllbpara\\
		&=\partIi+\partIii+\partIiii.
	\end{split}
\end{equation*}
For $\partIi$, by $\assStute$ and $\assEigen$(i), with propbability converging to 1, it holds that
$$
\partIi=\dAllbpara^\rmT\AllX^\rmT\Lmb\bW\Lmb\AllX\dAllbpara\rate^2\geq c_1M^2\rate^2.
$$
For $\partIii$, by $\assEigen$(ii), with propbability converging to 1, it holds that
$$
|\partIii|=\OP(1)\ratePropen^2\dAllbpara^\rmT\AllX^\rmT\bW\AllX\dAllbpara\rate^2=\OP(1)\ratePropen^2\rate^2M^2=\op(1)\rate^2M^2.
$$
For $\partIiii$, by $\assEigen$(ii), with propbability converging to 1, it holds that 
$$
|\partIiii|=\OP(1)2\ratePropen\dAllbpara^\rmT\AllX^\rmT\bW\AllX\dAllbpara\rate^2=\OP(1)\ratePropen\rate^2M^2=\op(1)\rate^2M^2.
$$
Thus, $\partI$ is dominated by \partIi.

Then we consider $\partII$. It can be deduced that 
\begin{equation*}
	\begin{split}
		\partII&=2\rate\dAllbpara\AllX^\rmT\left\{\Lmb\bW\bepsilon+\Lmb\bW(\Lmbhat-\Lmb)\AllX\Allbpara^*-\Lmb\bW\left(\bYhat-\bY\right)\right\}\\
		&+2\rate\dAllbpara\AllX^\rmT\left\{\left(\Lmb-\Lmbhat\right)\bW\bepsilon+\left(\Lmb-\Lmbhat\right)\bW\left(\Lmbhat-\Lmb\right)\AllX\Allbpara^*-\left(\Lmb-\Lmbhat\right)\bW\left(\bYhat-\bY\right)\right\}.
	\end{split}
\end{equation*}
By \assStute, we have $|\rate\dAllbpara\AllX^\rmT\Lmb\bW\bepsilon|=\rate M \alldim^{1/2}\numAll^{-1/2}\OP(1)=\rate^2M\OP(1)$. By \assStute and \assEigen, we have $|\rate\dAllbpara\AllX^\rmT\Lmb\bW\left(\Lmbhat-\Lmb\right)\AllX\Allbpara^*|=\rate \ratePropen\Vert\Allbpara^*\Vert_2M \OP(1)=\rate^2M\OP(1)$. Similarly, we can deduced that $|\rate\dAllbpara\AllX^\rmT\Lmb\bW\left(\bYhat-\bY\right)|=\rate \rateMu M \OP(1)=\rate^2M\OP(1)$, 
$$
|\rate\dAllbpara\AllX^\rmT\left(\Lmbhat-\Lmb\right)\bW\bepsilon|=\OP(1)\rate M \ratePropen\Vert\AllX^\rmT\bW\bepsilon\Vert_2=\OP(1)\rate M \ratePropen \sqrt{\frac{\alldim}{\numAll}}=\op(1)\rate^2 M,
$$
$$
|\rate\dAllbpara\AllX^\rmT\left(\Lmbhat-\Lmb\right)\bW\left(\Lmbhat-\Lmb\right)\AllX\Allbpara^*|=\OP(1)\rate M \ratePropen^2\Vert\Allbpara^*\Vert_2=\op(1)\rate^2 M,
$$
$$
|\rate\dAllbpara\AllX^\rmT\left(\Lmbhat-\Lmb\right)\bW\left(\bYhat-\bY\right)|=\OP(1)\rate\rateMu\ratePropen M\Vert\bW^{1/2}\AllX\Vert_2=\op(1)\rate^2M.
$$
Thus, $\partII$ is dominated by $\partIi$ too. 

Finally, we consider $\partIII$. It can be deduced that
\begin{equation*}
	\begin{split}
		\partIII&=\sum_{j=1}^{\alldim}\left\{\MCP(|\Allpara_j^*+\rate\dAllpara_j|;\tuneHTE)-\MCP(|\Allpara_j^*|;\tuneHTE)\right\}
		+\sum_{j=p+1}^{2\alldim}\left\{\MCP(|\Allpara_j^*+\rate\dAllpara_j|;\tuneHTE)-\MCP(|\Allpara_j^*|;\tuneUC)\right\}\\
		&=\partIIIi+\partIIIii.
	\end{split}
\end{equation*}
It can be seen that
\begin{equation*}
	\begin{split}
		\partIIIi&\geq\sum_{j\in\supp_1}\dMCP(|\HTEpara_j^*|;\tuneHTE)(|\HTEpara_j^*+\rate\Delta\HTEpara_j|-|\HTEpara_j^*|)+\frac{1}{2}\sum_{j\in\supp_1}\ddMCP(|\HTEpara_j^*|;\tuneHTE)(|\HTEpara_j^*+\rate\Delta\HTEpara_j|-|\HTEpara_j^*|)^2(1+o(1))\\
		&\geq -\sum_{j\in\supp_1}\dMCP(|\HTEpara_j^*|;\tuneHTE)\rate|\Delta\HTEpara_j|
		-\frac{1}{2}\sum_{j\in\supp_1}\ddMCP(|\HTEpara_j^*|;\tuneHTE)\rate^2|\Delta\HTEpara_j|^2(1+o(1))\\
		&\geq -\max\limits_{j\in\supp_1}\left\{\dMCP(|\HTEpara_j^*|;\tuneHTE)\right\}\sqrt{\realdimHTE}\rate M-\frac{1}{2}\max\limits_{j\in\supp_1}\left\{\ddMCP(|\HTEpara_j^*|;\tuneHTE)\right\}\rate^2M^2(1+o(1))\\
		&=-O(1)\rate^2M-o(1)\rate^2M^2.
	\end{split}
\end{equation*}
The last equation holds by $\assRate$(i)(ii). 
Similarly, we can obtain that 
$\partIIIii\geq-O(1)\rate^2M-o(1)\rate^2M^2$. 
Thus, $\partIII$ is dominated by $\partIi$ too. 

Now we can see that all terms are dominated by $\partIi$, that is, for large enough constant $M$, $\min\limits_{||\dAllbpara||_2=M}\loss_{\tuneHTE,\tuneUC}(\Allbpara^*+\rate\dAllbpara,\nuisanhat|\alldata)>\loss_{\tuneHTE,\tuneUC}(\Allbpara^*,\nuisanhat|\alldata)$ holds with probability converging to 1. This completes the proof.

\subsection*{Proof of Theorem \ref{thm2}}
For $1\leq j\leq \alldim$, if $\Allpara_j=0$, but $\Allparahat_j\neq 0$, then it holds that
$$
0=-\bU_j^\rmT\wh{\Lmb}\bW\left(\bYhat-\Lmbhat\bU\wh{\btheta}\right)+\dot{\rho}(|\wh{\theta}_j|)=\text{(I)}+\text{(II)}.
$$
It can be deduced that $\text{(I)}=\OP(\rate)$. By \assRho (ii),  $\text{(II)}/\tuneHTE>0$ with probability converging to 1. Thus 
$\text{(II)}>M\tuneHTE$, where $M>0$ is a constant. 
However, $\tuneHTE/\rate\rightarrow\infty$, as $\numAll\rightarrow \infty$, which comes to the contradictory.
For $\alldim+1\leq j\leq 2\alldim$, if $\Allpara_j=0$, but $\Allparahat_j\neq 0$, then it can be similarly deduced the contradictory since $\tuneUC/\rate\rightarrow\infty$, as $\numAll\rightarrow \infty$. 

\subsection*{Proof of Theorem \ref{thm3}}
Define the derivative for nuisance function
$$
\Gd[\dnuisance]=\partial_t\E\scorefb\left(\XSA,\failureT;\Allbpara_{\supp}^*,\nuisan^*+t\dnuisance\right)\Big|_{t=0}.
$$
Let $\bHessian:=\partial_ {\Allbpara_{\supp}}\E\scorefb(\XSA,\failureT;\Allbpara_{\supp},\nuisan^*)=\E\left\{(1-\propen^*(\XS)^2)\AllX\AllX^\rmT\right\}$.
By Taylor expansion, we have
\begin{equation*}
	\begin{split}
		&\E\scorefb\left(\XSA,\failureT;\Allbparahat_{\supp},\nuisanhat\right)-\E\scorefb\left(\XSA,\failureT;\Allbpara_{\supp}^*,\nuisan^*\right)\\
		&=\bHessian\left(\Allbparahat_{\supp}-\Allbpara^*_{\supp}\right)+\Gd[\nuisanhat-\nuisan^*]+\partial_t^2\E\scorefb\left(\XSA,\failureT;\Allbpara_{\supp}^*+t\left(\Allbparahat_{\supp}-\Allbpara_{\supp}^*\right),\nuisan^*+t\left(\nuisanhat-\nuisan^*\right)\right)\Big|_{t=\bar{t}}.
	\end{split}
\end{equation*}
For any measurable function $f$, we use the notation  $\wt{\E}_nf(\wt{\bZ},\wt{T})=\sum_{i=1}^{n}w_if(\wt{\bZ}_i,\wt{T}_i)$ for empirical expectation under Stute's measure. 
Similarly, we use the notation 
$\Gstute f(\wt{\bZ},\wt{T})=\sqrt{n}(\sum_{i=1}^{n}w_if(\wt{\bZ}_i,\wt{T}_i)-\E f(\wt{\bZ},\wt{T}))$ for empirical process under Stute's measure. 
For our model, we have
$$
\Estute\scorefb\left(\XSA,\failureT;\Allbparahat_{\setUC},\nuisanhat\right)+\partial_{\Allbpara_{\supp}}\rho\left(\Allbparahat_{\supp}\right)=\bzero,
$$
where  $\Estute\scorefb(\XSA,\failureT;\Allbpara_{\supp},\nuisan)=-\AllX^\rmT\Lmb\bW\bY+\AllX^\rmT\Lmb\bW\Lmb\AllX\Allbpara_{\supp}$.
By this equation and the result of Taylor expansion, it holds that
\begin{align*}
	\sqrt{\numAll}\bqT\bHessian\left(\Allbparahat_{\supp}-\Allbpara^*_{\supp}\right)&=\sqrt{\numAll}\bqT\Estute\scorefb\left(\XSA,\failureT;\Allbpara_{\supp}^*,\nuisan^*\right)-\sqrt{\numAll}\bqT\Gd\left[\nuisanhat-\nuisan^*\right]\\
	&\quad-\sqrt{\numAll}\bqT\partial_t^2\E\scorefb\left(\XSA,\failureT;\Allbpara_{\supp}^*+t\left(\Allbparahat_{\supp}-\Allbpara_{\supp}^*\right),\nuisan^*+t\left(\nuisanhat-\nuisan^*\right)\right)\Big|_{t=\bar{t}}\\
	&\quad+\bqT\Gstute\left\{\scorefb\left(\XSA,\failureT;\Allbparahat_{\supp},\nuisanhat\right)-\scorefb\left(\XSA,\failureT;\Allbpara_{\supp}^*,\nuisan^*\right)\right\}\\
	&\quad+ \sqrt{\numAll}\bqT\partial_{\Allbpara_{\supp}}\rho\left(\Allbparahat_{\supp}\right)\\
	&=\partI+\partII+\partIII+\partIV,
\end{align*}
where $\bar{t}\in(0,1)$.

First, we consider $\partI$. It can be deduced that $\Gd[\nuisanhat-\nuisan^*]=\bzero$. Then by $\assStute$ and $\sigma_*^2$ defined in the conditions, we have
$\partI\rightarrow_d N(0,\sigma_*^2)$. Then for $\partII$, it can be deduced that 
$$
\Vert\partial_t^2\E\scorefb\left(\XSA,\failureT;\Allbpara_{\supp}^*+t\left(\Allbparahat_{\supp}-\Allbpara_{\supp}^*\right),\nuisan^*+t\left(\nuisanhat-\nuisan^*\right)\right)|_{t=\bar{t}}\Vert_2=O(1)\rate^2.
$$
Then by $\assRate$(iv), we have $|\partII|\leq\sqrt{\numAll}\rate^2O(1)=o(1)$. For $\partIV$, we have
\begin{align*}
	|\partIV|&\leq \sqrt{\numAll}\Vert\partial_{\Allbpara_{\supp}}\rho(\Allbpara^*_{\supp})\Vert_2+\sqrt{\numAll}\Vert\partial_{\Allbpara_{\supp}}\rho(\Allbparahat_{\supp})-\partial_{\Allbpara_{\supp}}\rho(\Allbpara^*_{\supp})\Vert_2\\
	&\leq\sqrt{\numAll\realdimHTE\left(\max\limits_{j\in\supp_1}\left\{\dMCP(|\Allpara_j^*|;\tuneHTE)\right\}\right)^2+\numAll\realdimUC\left(\max\limits_{j\in\supp_2}\left\{\dMCP(|\Allpara_j^*|;\tuneUC)\right\}\right)^2}\\
	&\quad		+\OP(1)\sqrt{\numAll}\max\left\{\tuneHTE,\tuneUC\right\}\rate
\end{align*}
By $\assRate$(iii)(iv), it follows that $|\partIV|=o(1)$.
Now we consider $\partIII$. Define class
$$
\setF_2=\left\{\scoref_j\left(\cdot;\Allbpara_{\supp},\nuisanhat\right)-\scoref_j\left(\cdot;\Allbpara_{\supp},\nuisan^*\right): j\in\supp, \Vert\Allbpara_{\supp}-\Allbpara^*_{\supp}\Vert_2\leq c\rate\right\},
$$
where $c$ is some positve constant. Then we can see that 
$$
|\partIII|\leq \sqrt{\realdim}\sup_{f\in\setF_2}|\Gstute f|.
$$
By $\assRate$(v), we have $\sup_{f\in\setF_2}\Vert f\Vert^2_{\Fzt,2}\leq O(1)\rate^b$. Consider that
$$
\sup_{f\in\setF_2}|f|\leq \sup_{f\in\setF_{1,\nuisanhat}}|f|+\sup_{f\in\setF_{1,\nuisan^*}}|f|\leq \envelopF_{1,\nuisanhat}+\envelopF_{1,\nuisan^*},
$$
the function $\envelopF_2:=\envelopF_{1,\nuisanhat}+\envelopF_{1,\nuisan^*}$ can be the envelop for $\setF_2$. Since $\setF_2\subset \setF_{1,\nuisanhat}-\setF_{1,\nuisan^*}$, by $\assG$, it holds that
\begin{align*}
	&\sup_Q\log N(\xi\Vert\envelopF_2\Vert_{Q,2},\setF_2,\Vert\cdot\Vert_{Q,2})\\		
	&\leq\sup_Q\log N\left(\frac{\xi}{2}\Vert\envelopF_{1,\nuisanhat}\Vert_{Q,2},\setF_{1,\nuisanhat},\Vert\cdot\Vert_{Q,2}\right)
	+\sup_Q\log N\left(\frac{\xi}{2}\Vert\envelopF_{1,\nuisan^*}\Vert_{Q,2},\setF_{1,\nuisan^*},\Vert\cdot\Vert_{Q,2}\right)\\
	&\leq 2v\log\frac{2c_6}{\xi}.
\end{align*}
Under the condition $\assStute$, $\assRate$(v), and by Lemma 6.2 in \citet{Chernozhukov2018}, with probability converging to 1, for $q>2$, it holds that
$$
\sup_{f\in\setF_2}|\Gstute f|\leq 
O\left(\rate^{b/2}+\numAll^{-1/2+1/q}\right).
$$
Thus by $\assRate$(vi), we have $\partIII\leq \sqrt{\realdim}O(\rate^{b/2}+\numAll^{-1/2+1/q})=o(1)$.

Finally, combining the analysis of $\partI$ to $\partIV$, we have
$\sqrt{\numAll}\bqT\bHessian(\Allbparahat_{\supp}-\Allbpara^*_{\supp})\rightarrow_dN(0,\sigma_*^2)$. This completes the proof.

\subsection*{Proof of Theorem \ref{thm5}}
For abuse of notation, here we still use $\predictorX$ to denote $(\Treat-\propen(\XS))\predictorX$. 
Let 
\begin{align*}
	\bV_{rct}&=\E\left(\SetIndicator\predictorX_{\supp_1}\predictorX_{\supp_1}^\rmT\right),\\
	\bV_{int}&=\left[\E\predictorX_{\supp_1}\predictorX_{\supp_1}^\rmT-\left\{\E(1-\SetIndicator)\predictorX_{\supp_1}\predictorX_{\supp_2}^\rmT\right\}\left\{\E(1-\SetIndicator)\predictorX_{\supp_2}\predictorX_{\supp_2}^\rmT\right\}^{-1}\left\{\E(1-\SetIndicator)\predictorX_{\supp_2}\predictorX_{\supp_1}^\rmT\right\}\right].
\end{align*}
When there is no censoring, let $\bq\in \R^{\realdimHTE}$, it can be seen that as $\numAll\rightarrow\infty$,
\begin{align*}
	\var(\sqrt{\numAll}\bq^\rmT\HTEbparahat^{rct}_{\supp_1})&\rightarrow \frac{1}{\propS}\bq^\rmT\left\{\E\left(\predictorX_{\supp_1}\predictorX_{\supp_1}^\rmT|\SetIndicator=1\right)\right\}^{-1}\bq^\rmT=\bq^\rmT\bV_{rct}^{-1}\bq^\rmT,\\
	\var(\sqrt{\numAll}\bq^\rmT\HTEbparahat_{\supp_1})&\rightarrow \bq^\rmT\bV_{int}^{-1}\bq^\rmT.
\end{align*}
It can be deduced that
\begin{equation*}
	\begin{split}
		&\bV_{rct}-\bV_{int}\\
		&=\E(1-\SetIndicator)\predictorX_{\supp_1}\predictorX_{\supp_1}^\rmT-\left\{\E(1-\SetIndicator)\predictorX_{\supp_1}\predictorX_{\supp_2}^\rmT\right\}\left\{\E(1-\SetIndicator)\predictorX_{\supp_2}\predictorX_{\supp_2}^\rmT\right\}^{-1}\left\{\E(1-\SetIndicator)\predictorX_{\supp_2}\predictorX_{\supp_1}^\rmT\right\}. 
	\end{split}
\end{equation*}
By the theory of projection, $\bV_{rct}-\bV_{int}\geq 0$, where the equality holds if and only if there exists a $\realdimUC\times\realdimHTE$ constant matrix $\bQ$, such that when $S=0$, it holds that $\predictorX_{\supp_1}=\bQ^\rmT\predictorX_{\supp_2}$. 

As a result, 
\begin{equation}\label{variance_pf}
	\var(\sqrt{\numAll}\bq^\rmT\HTEbparahat_{\supp_1})^{-1}-\var(\sqrt{\numAll}\bq^\rmT\HTEbparahat^{rct}_{\supp_1})^{-1}\geq 0,
\end{equation} 
where the equality holds if and only if there exists a $\realdimUC\times\realdimHTE$ constant matrix $\bQ$, such that when $S=0$, it holds that $\predictorX_{\supp_1}=\bQ^\rmT\predictorX_{\supp_2}$. Specially, when $\supp_1\subset\supp_2$, the equality holds. When $\predictorX_{\supp_1}\perp\predictorX_{\supp_2}|S=0$ or $\supp_2=\emptyset$, under (B1)(i), 
it holds that  $\bV_{rct}-\bV_{int}=\E(1-S)\bX_{\mathcal{D}_1}\bX_{\mathcal{D}_1}^\rmT>0$, which means the ``$>$" in (\ref{variance_pf}) strictly holds.

\subsection*{Additional assumptions for Theorem \ref{thm4}}
We need some conditions similar to $\assStute$-$\assG$ to derive the asymptotic properties of the RCT-only estimator.
Denote that the true parameters for HTE be $\HTEbpara^*$, and true nuisance functions be $\nuisan^*$.
We first introduce some notations. Following the notations in \citet{Stute1996}, let $\Fcr$ be the probability distribution function (p.d.f) of $\Censore$ conditional on $\SetIndicator=1$, with
$\tau_{Gr}=\inf\left\{x:G_r(x)=1\right\}$, 
$F_r$ be the p.d.f of $\failureT$ conditional on $\SetIndicator=1$, with
$\tau_{Fr}=\inf\left\{x:F_r(x)=1\right\}$, and
$\Fyr$ be the p.d.f of $\obsY$  conditional on $S=1$, with $\Hendr=\inf\left\{x:\Fyr(x)=1\right\}$.
Let $\Fztr$ be the p.d.f of $(\XA,\failureT)$ conditional on $\SetIndicator=1$ with $\XA=\left(\predictorX,\Treat\right)$.  Define
\begin{equation*}
	\Ftildr(\bx,t)=\left\{
	\begin{array}{lc}
		\Fztr(\bx,t), & t<\Hendr,\\
		\Fztr(\bx,\Hendr-)+\Fztr(\bx,\Hendr)I(\Hendr\in\setH_r),& t\geq\Hendr,\\
	\end{array}
	\right.
\end{equation*}
with $\setH_r$ denoting the set of atoms of $\Fyr$, possibly empty. 
Define the score function
$$
\scorefr_j(\XA,\failureT;\HTEbpara^*,\nuisan^*)=\left\{A-\propen^*(\predictorX,1)\right\}\predictorX_j\left[\log\failureT-\MeanOutcome^*(\predictorX,1)-\left\{A-\propen^*(\predictorX,1)\right\}\predictorX^\rmT\HTEbpara^*\right],\quad j=1,2,...,\alldim.
$$
It can be seen that $\E\scorefbr(\XA,\failureT;\HTEbpara^*,\nuisan^*)=\bzero$.
Define 
$\Htild^{1}_r(\bx,y)=\Pr(\XA\leq \bx, \obsY\leq y, \CensoreIndicator=1|\SetIndicator=1)$, 
$\Htild^{0}_r(y)=\Pr(\obsY\leq y, \CensoreIndicator=0|\SetIndicator=1)$, and
$$
\stuteGam_{r0}(y)=\exp\left\{\int^{y-}_0\frac{\Htild^0_r(dv)}{1-\Fyr(v)}\right\},
$$
$$
\stuteGam_{r1j}(y)=\frac{1}{1-\Fyr(y)}\int 1_{\left\{y<v\right\}}\scorefr_j(\bz,v;\HTEbpara^*,\nuisan^*)\stuteGam_{r0}(v)\Htild^{1}_r(d\bz,dv),\quad j=1,2,...,\alldim,
$$
$$
\stuteGam_{r2j}(y)=\int\int \frac{1_{\left\{v<y,v<w\right\}}\scorefr_j(\bz,w;\HTEbpara^*,\nuisan^*)\stuteGam_{r0}(v)}{\left\{1-\Fyr(v)\right\}^2}\Htild_r^0(dv)\Htild^{1}_r(d\bz,dw),\quad j=1,2,...,\alldim.
$$
Let $\bstuteGam_{r1}(y)=\left\{\stuteGam_{r1j}(y),j=1,2,...,\alldim\right\}$, and $\bstuteGam_{r2}(y)=\left\{\stuteGam_{r2j}(y),j=1,2,...,\alldim\right\}$.

\begin{itemize}
	\item [\assStuter] $\Pr(\failureT\leq\Censore|\predictorX,\SetIndicator=1,\Treat,\failureT)=\Pr(\failureT\leq\Censore|\SetIndicator=1,\failureT)$.
	For $j=1,2,...,2\alldim$,
	
	(i) The p.d.f. $F_r$ and $G_r$ have no jump in common, and $\tau_{Fr}<\tau_{Gr}$.
	
	(ii) $\E\left\{\scorefr_j(\XSA,\obsY;\HTEbpara^*,\nuisan^*)\stuteGam_{r0}(\obsY)\CensoreIndicator|\SetIndicator=1\right\}^2<\infty$;
	
	(iii) Let $g_r(y)=\int_0^{y-}\left\{1-H_r(w)\right\}^{-1}\left\{1-G_r(w)\right\}^{-1}G_r(dw)$. It holds that
	$$\int|\scorefr_j(\bz,w;\HTEbpara^*,\nuisan^*)|\sqrt{g_r(w)}\Ftildr(d\bz,dw)<\infty$$.

	\item [\assEigenr]
	
	(i) The eigenvalues of $\E\left[\left\{\Treat-\propen(\predictorX,1)\right\}^2\predictorX\predictorX^\rmT|\SetIndicator=1\right]$ are larger than a positive constant $c_{r1}$.
	
	(ii) The eigenvalues of $\E\left(\predictorX\predictorX^\rmT|\SetIndicator=1\right)$ are smaller than a positive constant $c_{r2}$.
	
	\item [\assRater] Let $\ratePropen$ be the convergence rate of $\Vert\propenhat-\propen^*\Vert_\infty$, $\rateMu$ be the convergence rate of $\Vert\MeanOutcomehat-\MeanOutcome^*\Vert_\infty$. Define rate
	$\rater=\max\left\{\rateMu,\ratePropen\Vert\Allbpara^*\Vert_2,\sqrt{\alldim/\numAll}\right\}$. Let $\supp_1=\left\{j|\HTEpara^*_j\neq 0\right\}$ with element number $\realdimr$. Define $\dMCP(x;\lambda)=\partial\MCP(x;\lambda)/\partial x$, $\ddMCP(x;\lambda)=\partial^2\MCP(x;\lambda)/\partial x^2$, for $x>0$.	
	
	(i)
	$\sqrt{\realdimr}\max\left\{|\dMCP(|\HTEpara_j^*|;\tuneHTE)|,j=1,2,...,\alldim \right\}=\OP(\numRCT^{-1/2})$.
	
	(ii) $\sqrt{\numRCT}\tuneHTE\rater=o(1)$.
	
	(iii) $\E|\scorefr_j(\XA,\failureT;\HTEbpara,\nuisan)-\scorefr_j(\XA,\failureT;\HTEbpara^*,\nuisan^*)|^2\leq \left(\Vert\HTEbpara-\HTEbpara^*\Vert_2\land\Vert\nuisan-\nuisan^*\Vert_\infty\right)^{b_r}c_{r3}$, where $c_{r3}$ is some positive constant. In addition, $\sqrt{\realdimr}\rater^{b_r/2}=o(1)$, $\sqrt{\realdimr}\numRCT^{-1/2+1/q}=o(1)$, $q>2$.
	
	\item [\assGr] Let $\setHTEbpara=\left\{\HTEbpara:\Vert\HTEbpara-\HTEbpara^*\Vert_2\leq \rater c_{r4}\right\}$, where $c_{r4}$ is some positive constant. Define class 
	$$
	\setF_{r1,\nuisan}=\left\{\scorefr_j(\cdot;\HTEbpara,\nuisan):j\in\supp_1,\HTEbpara\in\setHTEbpara\right\},
	$$
	with measurable envelop $\envelopF_{r1,\nuisan}$. It satisfies  $\Vert\envelopF_{r1,\nuisan}\Vert_{\Fztr,q}\leq c_{r5}$ where $c_{r5}$ is some positive constant. It holds that for all $0<\xi\leq 1$,
	$$
	\sup_Q\log N(\xi\Vert\envelopF_{r1,\nuisan}\Vert_{Q,2},\setF_{r1,\nuisan},\Vert\cdot\Vert_{Q,2})\leq v\log\frac{c_{r6}}{\xi},
	$$
	where $c_{r6}$ is some positive constant.
\end{itemize}

\subsection*{Additional simulation experiments}

\begin{itemize}
	\item [1.] {\bf Weaker signal of the coefficients}: let $Signal=1$, and the other model settings remain the same as in Section 4 in the paper. The simulation results based on 500 replicates are shown in Table \ref{tabMSE1} and Table \ref{tabTIRFDR1}. The proposed method is still eﬀective. Compared with the results of $Signal=2$, it can be observed that when the signal of parameters get stronger, for all methods, RMSE, TIR, FDR improve.
	
	\item [2.] {\bf Logistic distribution of the error term}: let the error term follow the Logistic distribution instead of the Normal distribution, i.e., the survival time follows the log-Logistic distribution. We consider the case where the dimension of covariates is $p=20$; the censoring rate is CR$=40\%$ and $Signal=2$ for the proposed model and outcome adjusted model. The other model settings remain the same asin Section 4 of the paper. The simulation results based on 500 replicates are shown in Table \ref{logistic}. It can be seen that the proposed method still performs well and is more efficient than the outcome-adjusted method. Compared with the log-Normal distribution, all the methods lose some efficiency, i.e., higher RMSE.
	
	\item [3.] {\bf Severe censoring}: Let the censoring rate be CR$= 60\%$. We consider the case where the dimension of covariates is $p = 20$ and $Signal = 2$ for the proposed model and outcome-adjusted model. The other model settings remain the same as in Section 4 of the paper. The simulation results based on 500 replicates are shown in Table \ref{CR60}. It can be seen that the proposed method still works well and is more eﬃcient than the outcome-adjusted method. 
\end{itemize}

\begin{table}[h]\footnotesize
	\setstretch{1.2}
	\setlength{\tabcolsep}{2mm}
	\begin{threeparttable}
		\caption{The RMSE ($\times10^{2}$) of the HTE estimation when $Signal=1$ over 500 experiment replicates}\label{tabMSE1}
		\begin{tabular}{lcccccccc}
			\toprule
			\multicolumn{1}{c}{\multirow{3}{*}{Methods}} & \multicolumn{4}{c}{with unmeasured confounding}                                       & \multicolumn{4}{c}{no unmeasured confounding}                  \\
			\cmidrule(r){2-5}
			\cmidrule(r){6-9}
			\multicolumn{1}{c}{}                         & \multicolumn{2}{c}{p=20}                  & \multicolumn{2}{c}{p=50}                  & \multicolumn{2}{c}{p=20}                  & \multicolumn{2}{c}{p=50}                  \\
			\cmidrule(r){2-3}
			\cmidrule(r){4-5}
			\cmidrule(r){6-7}
			\cmidrule(r){8-9}
			\multicolumn{1}{c}{}                         & CR=20\%             & CR=40\%             & CR=20\%             & CR=40\%             & CR=20\%       & CR=40\%        & CR=20\%       & CR=40\%       \\
			\midrule
			\textbf{RL.or}                               & 6.54                & 7.52                & 4.15                & 4.97                & 4.59          & 5.37           & 2.96          & 3.42          \\
			\textbf{RL.RCTor}                            & 10.10               & 11.87               & 6.22                & 7.45                & 10.10         & 11.87          & 6.22          & 7.45          \\
			\textbf{RL.NAIor}                            & 35.75               & 35.96               & 22.54               & 22.72               &               &                &               &               \\
			\textbf{RL.cv}                               & {\uline {\textbf{7.06}}} & {\uline {\textbf{9.05}}} & {\uline {\textbf{5.41}}} & {\uline {\textbf{9.32}}} & \textbf{4.91} & \textbf{5.68}  & \textbf{3.08} & \textbf{3.53} \\
			\textbf{RL.bic}                              & 10.56               & 16.94               & 12.60               & 17.55               & \uline{4.68}    & \uline{5.47}     & \uline{2.99}    & \uline{3.46}    \\
			\textbf{RL.RCT}                              & 11.08               & 14.29               & 7.03                & 9.52                & 11.08         & 14.29          & 7.03          & 9.52          \\
			\textbf{RL.NAI}                              & 37.05               & 37.78               & 23.92               & 24.45               &               &                &               &               \\
			\midrule
			\textbf{OA.or}                               & 7.30                & 8.41                & 4.70                & 5.67                & 5.60          & 6.70           & 3.68          & 4.48          \\
			\textbf{OA.RCTor}                            & 12.43               & 14.60               & 8.70                & 10.18               & 12.43         & 14.60          & 8.70          & 10.18         \\
			\textbf{OA.cv}                               & \textbf{7.69}       & \textbf{9.67}       & \textbf{6.27}       & \textbf{9.44}       & \textbf{5.73} & \textbf{6.85}  & \textbf{3.73} & \textbf{4.56} \\
			\textbf{OA.bic}                              & 11.94               & 17.22               & 13.25               & 16.93               & 5.63          & 6.73           & 3.70          & 4.50          \\
			\textbf{OA.RCT}                              & 12.65               & 15.09               & 8.92                & 10.68               & 12.65         & 15.09          & 8.92          & 10.68         \\
			\midrule
			\textbf{GM0.or}                              & 8.71                & 10.18               & 5.82                & 6.96                & 6.46          & 7.50           & 4.18          & 4.98          \\
			\textbf{GM0.RCTor}                           & 13.97               & 16.39               & 9.64                & 11.47               & 13.97         & 16.39          & 9.62          & 11.48         \\
			\textbf{GM0.cv}                              & \textbf{9.29}       & \textbf{11.64}      & \textbf{7.04}       & \textbf{9.99}       & \textbf{7.03} & \textbf{8.37}  & \textbf{4.54} & \textbf{5.47} \\
			\textbf{GM0.biv}                             & 10.01               & 14.90               & 10.48               & 14.61               & 6.65          & 7.90           & 4.30          & 5.12          \\
			\textbf{GM0.RCT}                             & 14.33               & 17.55               & 10.61               & 13.99               & 14.33         & 17.55          & 10.56         & 14.00         \\
			\midrule
			\textbf{GM1.or}                              & 7.91                & 9.26                & 5.29                & 6.34                & 6.11          & 7.22           & 3.97          & 4.77          \\
			\textbf{GM1.RCTor}                           & 13.17               & 15.34               & 9.12                & 10.49               & 13.18         & 15.29          & 9.12          & 10.49         \\
			\textbf{GM1.cv}                              & \textbf{8.53}       & \textbf{10.69}      & \textbf{6.22}       & \textbf{9.20}       & \textbf{6.53} & \textbf{7.85}  & \textbf{4.26} & \textbf{5.17} \\
			\textbf{GM1.bic}                             & 9.03                & 13.42               & 9.33                & 14.27               & 6.34          & 7.42           & 4.07          & 4.85          \\
			\textbf{GM1.RCT}                             & 13.48               & 16.11               & 9.74                & 12.28               & 13.49         & 16.05          & 9.72          & 12.23         \\
			\midrule
			\textbf{Meta.or}                             & 7.44                & 8.72                & 5.00                & 5.98                & 5.92          & 6.93           & 3.84          & 4.58          \\
			\textbf{Meta.RCTor}                          & 12.64               & 14.70               & 8.80                & 10.17               & 12.65         & 14.68          & 8.79          & 10.17         \\
			\textbf{Meta.cv}                             & \textbf{7.81}       & \textbf{9.76}       & \textbf{5.82}       & \textbf{8.33}       & \textbf{6.21} & \textbf{7.38}  & \textbf{4.06} & \textbf{4.89} \\
			\textbf{Meta.bic}                            & 8.35                & 12.39               & 8.65                & 12.72               & 6.07          & 7.13           & 3.92          & 4.66          \\
			\textbf{Meta.RCT}                            & 12.89               & 15.46               & 9.49                & 12.05               & 12.90         & 15.44          & 9.48          & 12.04         \\
			\midrule
			\textbf{GM01.or}                             & 8.55                & 10.37               & 5.93                & 7.39                & 6.60          & 8.21           & 4.29          & 5.51          \\
			\textbf{GM01.RCTor}                          & 15.10               & 19.23               & 10.56               & 13.53               & 15.09         & 19.26          & 10.53         & 13.52         \\
			\textbf{GM01.cv}                             & \textbf{9.55}       & \textbf{12.21}      & \textbf{7.03}       & \textbf{9.87}       & \textbf{8.06} & \textbf{10.63} & \textbf{5.45} & \textbf{8.35} \\
			\textbf{GM01.bic}                            & 9.10                & 11.50               & 6.51                & 8.72                & 7.64          & 9.86           & 5.00          & 7.18          \\
			\textbf{GM01.RCT}                            & 15.39               & 19.71               & 10.74               & 14.14               & 15.39         & 19.73          & 10.71         & 14.11   \\
			\bottomrule     
		\end{tabular}
		\footnotesize{
			Some results are marked in bold to make it clear for readers to compare different methods. The results that behave the best, except for the Oracle estimates (i.e., the smallest RMSE), are marked with underlines. \\
				In the table, CR represents cencoring rate. 
				Among these methods, those with names starting with ``RL" indicate the proposed model. RL.cv and RL.bic represent the proposed estimates under CV and BIC criteria, respectively. RL.RCT represents the estimate merely based on RCT data. RL.NAI is the naive estimate that completely ignores unmeasured confounding effect. 
				RL.or, RL.RCTor, and RL.NAIor are the oracle estimates of the integrative, RCT-only, and naive analysis, respectively. 
				Other methods with names starting with ``OA'', ``GM0'', ``GM1'', ``Meta'' and ``GM01'' are introduced in detail in Section 4. 
		}
	\end{threeparttable}
\end{table}

\begin{table}[h]\footnotesize
	\setstretch{1.2}
	\setlength{\tabcolsep}{1.5mm}
	\begin{threeparttable}
		\caption{The averaged TIR($/\%$) and FDR($/\%$) when Signal=1 over 500 experiment replicates.}\label{tabTIRFDR1}
		\begin{tabular}{clcccccccc}
			\toprule
			\multirow{3}{*}{Index} & \multicolumn{1}{c}{\multirow{3}{*}{Methods}} & \multicolumn{4}{c}{with unmeasured confounding}     & \multicolumn{4}{c}{no unmeasured confounding}       \\
			\cmidrule(r){3-6}
			\cmidrule(r){7-10}
			& \multicolumn{1}{c}{}                         & \multicolumn{2}{c}{p=20} & \multicolumn{2}{c}{p=50} & \multicolumn{2}{c}{p=20} & \multicolumn{2}{c}{p=50} \\
			\cmidrule(r){3-4}
			\cmidrule(r){5-6}
			\cmidrule(r){7-8}
			\cmidrule(r){9-10}
			& \multicolumn{1}{c}{}                         & CR=20\%     & CR=40\%    & CR=20\%     & CR=40\%    & CR=20\%     & CR=40\%    & CR=20\%     & CR=40\%    \\
			\midrule
			\multirow{12}{*}{TIR}  & \textbf{RL.cv}                               & 100         & 100        & 100         & 100        & 93.00       & 92.80      & 94.00       & 90.80      \\
			& \textbf{RL.bic}                              & 100         & 100        & 100         & 99.00      & 97.40       & 96.40      & 95.40       & 94.40      \\
			\cmidrule{2-10}
			& \textbf{OA.cv}                               & 100         & 100        & 100         & 100        & 96.40       & 93.20      & 95.60       & 94.00      \\
			& \textbf{OA.bic}                              & 100         & 100        & 100         & 99.00      & 98.00       & 97.60      & 97.60       & 95.80      \\
			\cmidrule{2-10}
			& \textbf{GM0.cv}                              & 100         & 100        & 100         & 99.73      & 83.60       & 81.20      & 81.61       & 77.95      \\
			& \textbf{GM0.biv}                             & 100         & 100        & 100         & 99.47      & 89.20       & 88.20      & 92.41       & 88.98      \\
			\cmidrule{2-10}
			& \textbf{GM1.cv}                              & 100         & 100        & 100         & 100        & 85.20       & 80.60      & 82.53       & 77.95      \\
			& \textbf{GM1.bic}                             & 100         & 100        & 100         & 100        & 91.00       & 86.20      & 88.97       & 86.35      \\
			\cmidrule{2-10}
			& \textbf{Meta.cv}                             & 100         & 100        & 100         & 100        & 73.20       & 67.40      & 70.57       & 63.78      \\
			& \textbf{Meta.bic}                            & 100         & 100        & 100         & 100        & 83.40       & 77.20      & 82.76       & 78.74      \\
			\cmidrule{2-10}
			& \textbf{GM01.cv}                             & 100         & 100        & 100         & 100        & 17.20       & 6.20       & 9.20        & 0.26       \\
			& \textbf{GM01.bic}                            & 100         & 100        & 100         & 100        & 32.40       & 14.20      & 19.77       & 3.15       \\
			\midrule
			\multirow{19}{*}{FDR}  & \textbf{RL.cv}                               & 2.49        & 5.62       & 6.59        & 11.87      & 1.62        & 1.58       & 1.60        & 1.58       \\
			& \textbf{RL.bic}                              & 1.25        & 2.87       & 3.06        & 3.98       & 0.35        & 0.28       & 0.36        & 0.52       \\
			& \textbf{RL.RCT}                              & 6.39        & 11.16      & 8.38        & 15.55      & 6.39        & 11.16      & 8.38        & 15.55      \\
			& \textbf{RL.NAI}                              & 5.29        & 5.65       & 4.41        & 4.61       &             &            &             &            \\
			\cmidrule{2-10}
			& \textbf{OA.cv}                               & 2.16        & 4.96       & 5.19        & 8.40       & 0.94        & 1.20       & 1.00        & 1.65       \\
			& \textbf{OA.bic}                              & 1.34        & 2.11       & 1.60        & 2.41       & 0.13        & 0.13       & 0.35        & 0.31       \\
			& \textbf{OA.RCT}                              & 2.88        & 5.10       & 3.82        & 7.72       & 2.88        & 5.10       & 3.82        & 7.72       \\
			\cmidrule{2-10}
			& \textbf{GM0.cv}                              & 5.04        & 8.56       & 8.33        & 15.75      & 3.76        & 4.90       & 4.62        & 7.29       \\
			& \textbf{GM0.biv}                             & 2.32        & 3.47       & 4.04        & 6.57       & 0.90        & 1.32       & 0.63        & 1.69       \\
			& \textbf{GM0.RCT}                             & 2.88        & 5.02       & 2.57        & 6.15       & 2.88        & 5.02       & 2.54        & 6.09       \\
			\cmidrule{2-10}
			& \textbf{GM1.cv}                              & 3.86        & 8.11       & 7.35        & 15.23      & 3.00        & 4.28       & 4.47        & 6.29       \\
			& \textbf{GM1.bic}                             & 2.08        & 3.35       & 3.96        & 6.22       & 0.90        & 0.80       & 0.89        & 1.25       \\
			& \textbf{GM1.RCT}                             & 2.88        & 4.17       & 2.23        & 4.48       & 2.90        & 4.24       & 2.34        & 4.82       \\
			\cmidrule{2-10}
			& \textbf{Meta.cv}                             & 7.72        & 13.72      & 13.18       & 25.33      & 5.71        & 7.80       & 8.08        & 11.38      \\
			& \textbf{Meta.bic}                            & 3.87        & 5.95       & 7.02        & 11.14      & 1.64        & 2.01       & 1.39        & 2.71       \\
			& \textbf{Meta.RCT}                            & 5.51        & 8.38       & 4.61        & 9.92       & 5.48        & 8.42       & 4.64        & 10.18      \\
			\cmidrule{2-10}
			& \textbf{GM01.cv}                             & 7.84        & 16.61      & 20.29       & 41.2       & 6.53        & 15.23      & 15.73       & 36.5       \\
			& \textbf{GM01.bic}                            & 4.36        & 9.37       & 11.26       & 24.60      & 4.10        & 9.11       & 9.46        & 24.07      \\
			& \textbf{GM01.RCT}                            & 5.34        & 9.01       & 3.81        & 9.09       & 5.49        & 8.81       & 3.74        & 8.56 \\
			\bottomrule
		\end{tabular}
		\footnotesize{			
				In the table, CR represents the cencoring rate. TIR is the rate of correctly identifying the real case where unmeasured confounding effect exists or not. FDR is the false discovery rate of the HTE estimates. 
				Among these methods, those with names starting with ``RL" indicate the proposed model. RL.cv and RL.bic represent the proposed estimates under CV and BIC criteria, respectively. RL.RCT represents the estimate merely based on RCT data. RL.NAI is the naive estimate that completely ignores unmeasured confounding effect. 
				Other methods with names starting with ``OA'', ``GM0'', ``GM1'', ``Meta'' and ``GM01'' are introduced in detail in Section 4. 
		}
	\end{threeparttable}
\end{table}

\begin{table}[htpb]
	\setstretch{1.35}
	\setlength{\tabcolsep}{3.4mm}
	\caption{Simulation results of the proposed method (RL) and the outcome-adjusted method (OA) when the survival time follows log-Logistic distribution, CR=40\%, $p=20$. }\label{logistic}
	\vspace{0.1in}
	\begin{tabular}{clrrclrr}
		\toprule
		\textbf{Index}        & \textbf{Method} & \textbf{$\bbeta^*\neq\bzero$} & \textbf{$\bbeta^*=\bzero$}          & \textbf{Index}       & \textbf{Method} & \textbf{$\bbeta^*\neq\bzero$} & \textbf{$\bbeta^*=\bzero$}         \\
		\midrule
		\multirow{12}{*}{{RMSE}$\times 10^2$} & RL.or           & 14.81                          & 10.41        & \multirow{7}{*}{FDR/\%} & RL.cv           & 5.24                           & 1.65                 \\
		& RL.RCTor        & 22.56                          & 22.56        &                      & RL.bic          & 3.43                           & 0.37                 \\
		& RL.NAIor        & 71.86                          & -            &                      & RL.RCT          & 8.90                           & 8.90                 \\
		& RL.cv           & 17.74                          & 11.09        &                      & RL.NAI          & 6.69                           & 2.61                 \\
		\cmidrule(r){6-8}
		& RL.bic          & 27.01                          & 10.55        &                      & OA.cv           & 4.02                           & 1.39                 \\
		& RL.RCT          & 25.27                          & 25.27        &                      & OA.bic          & 1.95                           & 0.33                 \\
		& RL.NAI          & 75.16                          & -            &                      & OA.RCT          & 4.51                           & 4.50                 \\
		\cmidrule(r){2-4}\cmidrule(r){5-8}
		& OA.or           & 16.29                          & 13.19        & \multirow{5}{*}{TIR/\%} & RL.cv           & 100.00                         & 91.00                \\
		& OA.RCTor        & 28.91                          & 28.95        &                      & RL.bic          & 100.00                         & 93.40                \\
		\cmidrule(r){6-8}
		& OA.cv           & 18.03                          & 13.43        &                      & OA.cv           & 100.00                         & 93.81                \\
		& OA.bic          & 28.97                          & 13.24        &                      & OA.bic          & 100.00                         & 95.58                \\
		& OA.RCT          & 29.63                          & 29.67        &                      &                 & \multicolumn{1}{l}{}           & \multicolumn{1}{l}{}\\
		\bottomrule
	\end{tabular}
	
	\vspace{0.1in}
	\footnotesize{In the table, the results are based on 500 replicates. CR represents the cencoring rate. $\bbeta^*\neq\bzero$ means there is unmeasured confounding effect, otherwise, there is no unmeasured confounding effect. TIR is the rate of correctly identifying the real case where unmeasured confounding effect exists or not. FDR is the false discovery rate of the HTE estimates. 
		RL.cv and RL.bic represent the proposed estimates under the CV and BIC criteria, respectively. RL.RCT represents the estimate merely based on RCT data. RL.NAI is the naive estimate that completely ignores unmeasured confounding effect.}
\end{table}

\begin{table}[htpb]
	\setstretch{1.35}
	\setlength{\tabcolsep}{3.4mm}
	\caption{Simulation results of the proposed method (RL) and the outcome-adjusted method (OA) when CR=60\%, $p=20$.}\label{CR60}
	\vspace{0.1in}
	\begin{tabular}{clrrclrr}
		\toprule
		\textbf{Index}        & \textbf{Method} & \textbf{$\bbeta^*\neq\bzero$} & \textbf{$\bbeta^*=\bzero$}          & \textbf{Index}       & \textbf{Method} & \textbf{$\bbeta^*\neq\bzero$} & \textbf{$\bbeta^*=\bzero$}         \\
		\midrule
		\multirow{12}{*}{RMSE$\times 10^2$} & RL.or           & 14.81                          & 9.13                  & \multirow{7}{*}{FDR/\%} & RL.cv           & 3.60                           & 1.49                 \\
		& RL.RCTor        & 24.20                          & 24.20                 &                      & RL.bic          & 2.28                           & 0.44                 \\
		& RL.NAIor        & 71.46                          & - &                      & RL.RCT          & 9.75                           & 9.75                 \\
		& RL.cv           & 16.80                          & 9.74                  &                      & RL.NAI          & 6.31                           & 1.96                 \\
		\cmidrule(r){6-8}
		& RL.bic          & 20.77                          & 9.33                  &                      & OA.cv           & 1.31                           & 0.53                 \\
		& RL.RCT          & 27.71                          & 27.71                 &                      & OA.bic          & 0.54                           & 0.19                 \\
		& RL.NAI          & 74.23                          & - &                      & OA.RCT          & 3.45                           & 3.45                 \\
		\cmidrule(r){2-4}\cmidrule(r){5-8}
		& OA.or           & 17.08                          & 14.27                 & \multirow{5}{*}{TIR/\%} & RL.cv           & 100.00                         & 94.80                \\
		& OA.RCTor        & 32.38                          & 32.36                 &                      & RL.bic          & 100.00                         & 95.20                \\
		\cmidrule(r){6-8}
		& OA.cv           & 17.56                          & 14.38                 &                      & OA.cv           & 100.00                         & 97.74                \\
		& OA.bic          & 18.69                          & 14.31                 &                      & OA.bic          & 100.00                         & 98.02                \\
		& OA.RCT          & 32.95                          & 32.95                 &                      &                 & \multicolumn{1}{l}{}           & \multicolumn{1}{l}{}\\
		\bottomrule
	\end{tabular}
	
	\vspace{0.1in}
	\footnotesize{
		In the table, the results are based on 500 replicates.  CR represents the cencoring rate. $\bbeta^*\neq\bzero$ means there is unmeasured confounding effect, otherwise, there is no unmeasured confounding effect. TIR is the rate of correctly identifying the real case where unmeasured confounding effect exists or not. FDR is the false discovery rate of the HTE estimates. 
		RL.cv and RL.bic represent the proposed estimates under the CV and BIC criteria, respectively. RL.RCT represents the estimate merely based on RCT data. RL.NAI is the naive estimate that completely ignores unmeasured confounding effect.}
\end{table}


\begin{thebibliography}{}
	%
	\bibitem[{Altorki et~al.(2023)}]{Altorki2023}
	{Altorki, N., Wang, X., Kozono, D., Watt, C., et al.},
	\newblock {Lobar or sublobar resection for peripheral stage IA non–small-cell lung cancer},
	\newblock \textit{New England Journal of Medicine},
	\textbf{388},
	489--498 (2023)
	
	\bibitem[{Angrist et~al.(1996)}]{Angrist1996}
	{Angrist, J., Imbens, G. and Rubin, D.},
	\newblock {Identification of causal effects using instrumental variables},
	\newblock \textit{Journal of the American Statistical Association},
	\textbf{91},
	468--472 (1996)
	
	\bibitem[{Breiman(2001)}]{Breiman2001}
	{Breiman, L.},
	\newblock {Random forests},
	\newblock \textit{Machine Learning},
	\textbf{45},
	5--32 (2001)
	
	\bibitem[{Cheng et al.(2023)}]{Cheng2023}
	{Cheng, Y., Wu, L. and Yang, S.},
	\newblock {Enhancing treatment effect estimation: A model robust approach integrating randomized experiments and external controls using the double penalty integration estimator},
	\newblock \textit{39th Conference on Uncertainty in Artificial Intelligence (UAI 2023)},
	accepted (2023)
	
	
	\bibitem[{Chernozhukov et~al.(2018)}]{Chernozhukov2018}
	{Chernozhukov, V., Chetverikov, D., Demirer, M, Duflo, E., et al.},
	\newblock {Double/debiased machine learning for treatment and structural parameters},
	\newblock \textit{Econometrics Journal},
	\textbf{21},
	1--68 (2018)
	
	\bibitem[{Collins and Varmus(2015)}]{Collins2015}
	{Collins, F. and Varmus, H.},
	\newblock {A new initiative on precision medicine},
	\newblock \textit{New England Journal of Medicine},
	\textbf{372},
	793--795 (2015)
	
	\bibitem[{Efron and Tibshirani(1993)}]{Efron1993}
	{Efron, B. and Tibshirani, R.},
	\newblock {An introduction to the bootstrap},
	\newblock \textit{New York: Chapman and Hall} (1993)
	
	
	\bibitem[{Fan and Peng(2004)}]{Fan2004}
	{Fan, J. and Peng, H.},
	\newblock {Nonconcave penalized likelihood with a diverging number of parameters},
	\newblock \textit{The Annals of Statistics},
	\textbf{32},
	928--961 (2004)
	
	\bibitem[{Ginsberg and Rubinstein(1995)}]{Ginsberg1995}
	{Ginsberg, R. and Rubinstein, L.},
	\newblock {Randomized trial of lobectomy versus limited resection for T1 N0 non-small cell lung cancer. Lung Cancer Study Group},
	\newblock \textit{Journal of the American Statistical Association},
	\textbf{60},
	615--623 (1995)
	
	
	\bibitem[{Guo et~al.(2021)}]{Guo2021}
	{Guo, W., Zhou, X. and Ma, S.},
	\newblock {Estimation of optimal individualized treatment rules Using a covariate-specific treatment effect curve with high-dimensional covariates},
	\newblock \textit{Journal of the American Statistical Association},
	\textbf{116},
	309--321 (2021)
	
	
	\bibitem[{Hamburg and Collins(2010)}]{Hamburg2010}
	{Hamburg, M. A. and Collins, F. S. }, 
	\newblock {The path to personalized medicine},
	\newblock \textit{New England Journal of Medicine},
	\textbf{363},
	301--304 (2010)
	
	\bibitem[{He et~al.(2016)}]{He2016}
	{He Q., Zhang H. H., Avery C. L., and Lin D.},
	\newblock {Sparse meta-analysis with high-dimensional data},
	\newblock \textit{Biostatistics},
	\textbf{17},
	205--220 (2016)
	
	\bibitem[{Henderson et~al.(2020)}]{Henderson2020}
	{Henderson, N., Louis, T., Rosner, G, and Varadhan, R.},
	\newblock {Individualized treatment effects with censored data via fully nonparametric Bayesian accelerated failure time models},
	\newblock \textit{Biostatistics},
	\textbf{21},
	50--68 (2020)
	
	\bibitem[{Hu et~al.(2021)}]{Hu2021}
	{Hu, L., Ji, J., and Li, F.},
	\newblock {Estimating heterogeneous survival treatment effect in observational data using machine learning},
	\newblock \textit{Statistics in Medicine},
	\textbf{40},
	4691--4713 (2021)
	
	\bibitem[{Huang et~al.(2006)}]{Huang2006}
	{Huang, J., Ma, S., and Xie, H.},
	\newblock {Regularized estimation in the accelerated failure time model with high-dimensional covariates},
	\newblock \textit{Biometrics},
	\textbf{62},
	813--820 (2006).
	
	\bibitem[{Kallus et~al.(2018)}]{Kallus2018}
	{Kallus, N., Puli, A., and Shalit, U.},
	\newblock {Removing hidden confounding by experimental grounding},
	\newblock \textit{Advances in Neural Information Processing Systems},
	\textbf{31},
	10911--10920 (2018)
	
	\bibitem[{Kuroki and Pearl(2014)}]{Kuroki2014}
	{Kuroki, M. and Pearl, J.},
	\newblock {Measurement bias and effect restoration in causal inference},
	\newblock \textit{Biometrika},
	\textbf{101},
	423--437 (2014)
	
	\bibitem[{Kunzel et~al.(2019)}]{Kunzel2019}
	{Kunzel, S., Sekhon, J., Bickel, P., et al.},
	\newblock {Metalearners for estimating heterogeneous treatment effects using machine learning},
	\newblock \textit{Proceedings of the National Academy of Sciences of the United States of America},
	\textbf{116},
	4156--4165 (2019)
	
	\bibitem[{Lee and Altorki(2023)}]{Lee2023}
	{Lee, B. and Altorki, N.},
	\newblock {Sub-lobar resection: the new standard of care for early-stage lung cancer},
	\newblock \textit{Cancers},
	\textbf{15},
	2914 (2023)
	
	\bibitem[{Lee et~al.(2022)}]{Lee2022}
	{Lee, D., Yang, S., and Wang, X.},
	\newblock {Doubly robust estimators for generalizing treatment effects on survival outcomes from randomized controlled trials to a target population},
	\newblock \textit{Journal of Causal Inference},
	\textbf{10},
	415--440. (2022)
	
	\bibitem[{Lee et~al.(2024)}]{Lee2024}
	{Lee, D., Gao, C., Ghosh, S., and Yang, S.},
	\newblock {Transporting survival of an HIV clinical trial to the external target populations},
	\newblock \textit{Journal of Biopharmaceutical Statistics},
	doi.org/10.1080/10543406.2024.2330216. (2024)
	
	
	
	\bibitem[{Ma and Zhou(2018)}]{Ma2018}
	{Ma Y. and Zhou X.},
	\newblock {Treatment selection in a randomized clinical trial via covariate-specific treatment effect curves},
	\newblock \textit{Statistical Methods in Medical Reasearch},
	\textbf{26},
	124--141 (2018)
	
	\bibitem[{Neyman(1959)}]{Neyman1959}
	{Neyman J.},
	\newblock {Optimal asymptotic tests of composite statistical hypotheses},
	\newblock \textit{Probability and Statsitics},
	213--234 (1959)
	
	\bibitem[{Nie and Wager(2018)}]{Nie2018}
	{Nie X. and Wager S.},
	\newblock {Quasi-oracle estimation of heterogeneous treatment effects},
	\newblock \textit{Biometrika},
	\textbf{108},
	299--319 (2021)
	
	
	\bibitem[{Prentice et~al.(2008)}]{Prentice2008}
	{Prentice, R. L., Chlebowski, R. T., Stefanick, M. L., et al.}, 
	\newblock {Estrogen plus progestin therapy and breast cancer in recently postmenopausal women},
	\newblock \textit{American Journal of Eepdemiology},
	\textbf{167},
	1207--1216 (2008)
	
	\bibitem[{Powers et~al.(2017)}]{Powers2017}
	{Powers, S., Qian J., Jung K., et al.},
	\newblock {Some methods for heterogeneous treatment effect estimation in high dimensions},
	\newblock \textit{Statistics in Medicine},
	\textbf{37},
	1767--1787 (2017)
	
	
	\bibitem[{Robins et~al.(1999)}]{Robins1999}
	{Robins, J. M., Rotnitzky, A. and Scharfstein, D. O.}, 
	\newblock {Sensitivity analysis for selection bias and unmeasured confounding in missing data and causal inference models},
	\newblock \textit{Statistical Models in Epidemiology, the  Environment, and Clinical Trials, Springer, New York},
	1--94 (1999)
	
	\bibitem[{Robinson(1988)}]{Robinson1988}
	{Robinson, P.}, 
	\newblock {Root-$n$-consistent semiparametric regression},
	\newblock \textit{Econometrica},
	\textbf{56},
	931--954 (1988)
	
	\bibitem[{Saji et~al.(2022)}]{Saji2022}
	{Saji, H.,  Okada, M., Tsuboi, M., Nakajima, R., et al.}, 
	\newblock {Segmentectomy versus lobectomy in small-sized peripheral non-small-cell lung cancer (JCOG0802/WJOG4607L): a multicentre, open-label, phase 3, randomised, controlled, non-inferiority trial},
	\newblock \textit{Lancet},
	\textbf{399},
	1607--1617 (2022)
	
	\bibitem[{Shalit et~al.(2017)}]{Shalit2017}
	{Shalit, U., Johansson, F., and Sontag, D.}, 
	\newblock {Estimating individual treatment effect: generalization bounds and algorithms},
	\newblock \textit{In Proceedings of the 34th International Conference on Machine Learning},
	\textbf{70},
	3076--3085 (2017)
	
	\bibitem[{Simoneau et~al.(2020)}]{Simoneau2020}
	{Simoneau, G., Moodie, E., Nijjar, J., and Platt, R.}, 
	\newblock {Estimating optimal dynamic treatment regimes with survival outcomes},
	\newblock \textit{Journal of the American Statistical Association},
	\textbf{115(531)},
	1531--1539 (2020)
	
	
	
	\bibitem[{Stute(1993)}]{Stute1993}
	{Stute, W.},
	\newblock {Consistent estimation under random censorship when covariates are present},
	\newblock \textit{Journal of Multivariate Analysis},
	\textbf{45},
	89--103 (1993)
	
	\bibitem[{Stute(1996)}]{Stute1996}
	{Stute, W.}, 
	\newblock {Distributional convergence under random censorship when covariables are present},
	\newblock \textit{Scandinavian Journal of Statistics},
	\textbf{23},
	461--471 (1996)
	
	\bibitem[{Sung et~al.(2021)}]{Sung2021}
	{Sung, H., Ferlay, J., Siegel, R., et al.}, 
	\newblock {Global cancer statistics 2020: GLOBOCAN estimates of incidence and mortality worldwide for 36 cancers in 185 countries},
	\newblock \textit{CA-A Cancer Journal for Clinicians},
	\textbf{71},
	209--249 (2021)
	
	
	\bibitem[{Verde and Ohmann(2015)}]{Verde2015}
	{Verde, P. E. and Ohmann, C.},
	\newblock {Combining randomized and non-randomized evidence in clinical research: a review of methods and applications},
	\newblock \textit{Research Synthesis Methods},
	\textbf{6},
	45--62 (2015)
	
	
	\bibitem[{Wager and Athey(2018)}]{Wager2018}
	{Wager S. and Athey S.},
	\newblock \textit{Estimation and inference of heterogeneous treatment effects using random forests},
	\newblock {Journal of the American Statistical Association},
	\textbf{113(523)},
	1228--1242 (2018)
	
	
	\bibitem[{Wang et~al.(2013)}]{Wang2013}
	{Wang L., Kim Y., and Li R.}, 
	\newblock \textit{Calibrating nonconvex penalized regression in ultra-high dimension},
	\newblock {The Annals of Statistics},
	\textbf{41(5)},
	2505--2536 (2013)
	
	\bibitem[{Wendling et~al.(2017)}]{Wendling2017}
	{Wendling, T., Jung, K., Callahan, A., et al.},
	\newblock {Comparing methods for estimation of heterogeneous treatment effects using observational data from health care databases},
	\newblock \textit{Statistics in Medicine},
	\textbf{37},
	3309--3324 (2017)
	
	\bibitem[{Wu and Yang(2022)}]{Wu22}
	{Wu, L. and Yang, S.},
	\newblock {Integrative R-learner of heterogeneous treatment effects combining
		experimental and observational studies},
	\newblock \textit{Proceedings of Machine Learning Research},
	\textbf{140},
	1--S5 (2022)
	
	\bibitem[{Yang et~al.(2020)}]{YangAFT2020}
	{Yang, S., Pieper, K. and Cools, F.},
	\newblock {Semiparametric estimation of structural failure time models in continuous-time processes},
	\newblock \textit{Biometrika},
	\textbf{107},
	123--136 (2020a)
	
	
	\bibitem[{Yang et~al.(2020)}]{Yang2020}
	{Yang, S., Zeng, D. and Wang, X.},
	\newblock {Improved inference for heterogeneous treatment effects using real-world data subject to hidden confounding},
	\newblock \textit{arXiv:2007.12922.} (2020b)
	
	
	\bibitem[{Yang et~al.(2020)}]{YangTest2020}
	{Yang, S., Zeng, D. and Wang, X.}, 
	\newblock {Elastic Integrative analysis of randomised trial and real-world data for treatment heterogeneity estimation},
	\newblock \textit{Journal of the Royal Statistical Society: Series B},
	\textbf{85},
	575--596 (2023)
	
	\bibitem[{Zhang(2010)}]{Zhang2010}
	{Zhang, C.}, 
	\newblock {Nearly unbiased variable selection under minimax concave penalty},
	\newblock \textit{The Annals of Statistics},
	\textbf{38},
	894--942 (2010)
	
	\bibitem[{Zhang et~al.(2019)}]{Zhang2019}
	{Zhang, Z., Feng, H., Zhao, H., Hu, J., et al.}, 
	\newblock {Sublobar resection is associated with better perioperative outcomes in elderly patients with clinical stage I non-small cell lung cancer: a multicenter retrospective cohort study},
	\newblock \textit{Journal of Thoracic Disease},
	\textbf{11},
	1838--1848 (2019)
	
	
	\bibitem[{Zhou and Zhu (2010)}]{Zhou2010}
	{Zhou, N. and Zhu, J.},
	\newblock {Group variable selection via a hierarchical lasso and its oracle property},
	\newblock \textit{Statistics and Its Inference},
	\textbf{3},
	557--574 (2010)
	
	\bibitem[{Zhou and Zhu (2021)}]{Zhou2021}
	{Zhou, N. and Zhu, L.},
	\newblock {On IPW-based estimation of conditional average treatment effects},
	\newblock \textit{Journal of Statistical Planning and Inference},
	\textbf{215},
	1--22 (2021)
	
	\bibitem[{Zhu and Gallego(2020)}]{Zhu2020}
	{Zhu, J. and Gallego, B.},
	\newblock {Targeted estimation of heterogeneous treatment effect in observational survival analysis},
	\newblock \textit{Journal of Biomedical Informatics},
	\textbf{107} (2020)
	
	\bibitem[{Zou(2006)}]{Zou2006}
	{Zou, H.}, 
	\newblock {The adaptive Lasso and its oracle properties},
	\newblock \textit{Journal of the American Statistical Association},
	\textbf{101}, 1418--1429 (2006)
	
	
	
\end{thebibliography}
\end{document}